\documentclass{aa}
\usepackage{supertabular}
\usepackage{psfig}
\usepackage{lscape}

\newcommand\ns{@{\hspace{0mm}}}
\newcommand\hs{@{\hspace{1.5mm}}}
\newcommand\Hs{@{\hspace{3.0mm}}}

\begin{document}

\title{The second ROSAT PSPC survey of \object{M31} and the complete ROSAT PSPC source list}

\author{R. Supper \inst{1} \and
		  G. Hasinger \inst{1,2} \and
		  W.H.G. Lewin \inst{3} \and
		  E.A. Magnier \inst{4} \and
		  J. van Paradijs$^\dagger$ \inst{5,6} \and
		  W. Pietsch \inst{1} \and
                  A.M. Read \inst{1} \and
		  J. Tr\"umper \inst{1}}

\institute{Max-Planck-Institut f\"ur extraterrestrische Physik, Postfach 1603, 85740 Garching, Germany \and
	   Astrophysikalisches Institut, An der Sternwarte 16, 14482 Potsdam, Germany \and
	   CSR and Department of Physics, Massachusetts Institute of Technology, Room 37-627, Cambridge, MA 02139, USA \and
           Canada-France-Hawaii Telescope, 65-1238 Mamalahoa Hwy, Kamuela, HI 96743, USA \and
	   Astronomical Institute ``Anton Pannekoek'', University of Amsterdam, Kruislaan 403, 1098 SJ Amsterdam, Netherlands \and
	   Physics Department, University of Alabama in Huntsville, Huntsville, AL 35899, USA }

\offprints{R. Supper, e-mail:ros@mpe.mpg.de
           $\dagger$Jan van Paradijs passed away on November 2nd, 1999.}

\date{Received date / Accepted date}

\abstract{
This paper reports the results of the analysis of the second ROSAT PSPC survey
of \object{M31} performed in summer 1992.  We compare our results with those 
of the first survey, already published in Supper et al.\
(\cite{Sup97}). Within the $\sim10.7 \mbox{ deg}^2$ field of view, 396
individual X-ray sources are detected in the second survey data, of
which 164 are new detections. When combined with the first survey,
this result in a total of 560 X-ray sources in the field of M31.
Their (0.1 keV -- 2.0 keV) fluxes range from $7 \times
10^{-15} \mbox{ erg cm}^{-2} \mbox{ s}^{-1}$ to $7.6 \times 10^{-12} \mbox{
erg cm}^{-2} \mbox{ s}^{-1}$, and of these 560 sources, 55 are tentatively
identified with foreground stars, 33 with globular clusters, 16 with supernova
remnants, and 10 with radio sources and galaxies (including \object{M32}). A
comparison with the results of the {\it Einstein} \object{M31} survey reveals
491 newly detected sources, 11 long term variable sources, and 7 possible transient
sources. Comparing the two ROSAT surveys, we come up with 34 long term
variable sources and 8 transient candidates. For the \object{M31} sources, the
observed luminosities range from $4 \times 10^{35} \mbox{ erg s}^{-1}$ to $4
\times 10^{38} \mbox{ erg s}^{-1}$. The total (0.1 keV -- 2.0 keV) luminosity
of \object{M31} is $(3.4\pm0.3) \times 10^{39} \mbox{ erg s}^{-1}$,
distributed approximately equally between the bulge and disk. Within the bulge
region, the luminosity of a possible diffuse component combined with faint
sources below the detection threshold is $(2.0\pm0.5) \times 10^{38} \mbox{
erg s}^{-1}$. An explanation in terms of hot gaseous emission leads to a
maximum total gas mass of $(1.0\pm0.3) \times 10^6 \, \mbox{M}_{\sun}$.
}

\maketitle

\section{Introduction}

Before ROSAT, the knowledge of the X-ray properties of \object{M31} was mainly
based on the IPC and HRI observations with the {\it Einstein} observatory.
These observations were performed in the years 1979 \& 1980 and the main
results are described in van Speybroeck et al.\ (\cite{Spe79}), van Speybroeck
\& Bechtold (\cite{Spe81}), Long \& van Speybroeck (\cite{Lon83}), Crampton
et al.\ (\cite{Cra84}), and Trinchieri \& Fabbiano (\cite{Tri91}, hereafter
TF). In 300 ks of total exposure, $\sim86$\% of the area of our neighbouring
spiral galaxy \object{M31} had been covered to a limiting sensitivity of
$\sim10^{37} \mbox{erg s}^{-1}$. Many of the 108 detected point sources within
these {\it Einstein} observations were measured with a positional accuracy
of about $3\arcsec$, and were found to be concentrated within a highly
confused bulge region and an outer region approximately following the spiral
arms. In addition, a variety of possible optical counterparts had been
determined, dividing into groups of foreground stars within our own Galaxy,
accreting objects and supernova remnants in \object{M31} and background
galaxies seen through the disc of \object{M31}. Additionally, it had been
considered that the luminosity distribution of the \object{M31} disk sources
were comparable with that of the bulge sources. The high confusion in the
bulge region together with the moderate total number of detected sources made
it difficult to justify this statement.

Two deep PSPC surveys of \object{M31} were performed with the ROSAT X-ray
observatory, the first in the summer of 1991, the second in the summer of 1992
(with some follow-up observations in the winter of 1992/1993). Both surveys
consisted of a number of observations arranged in raster pointings, covering the
whole area of \object{M31} and beyond. The total observation time of 200 ks
for each survey, together with the higher sensitivity of the ROSAT
telescope pushed the limiting flux to a factor of 10  -- 100 lower than
for the {\it Einstein} observations. Additionally, several ROSAT HRI M31 pointings were performed,
including a very deep one on the bulge. These were discussed by Primini et
al.\ (\cite{Pri93}) and Immler (\cite{Imm00}).

The results of the first PSPC survey have been described in Supper et al.\
(\cite{Sup97}, hereafter S97). This work led to the detection of 396 X-ray
point sources with (0.1 keV - 2.4 keV) fluxes ranging from $\sim5 \times
10^{-15} \mbox{erg cm}^{-2} \mbox{ s}^{-1}$ to $\sim4 \times 10^{-12}
\mbox{erg cm}^{-2} \mbox{ s}^{-1}$. Several tentative identifications with
foreground stars, globular clusters, supernova remnants, and 
galaxies were found, but the majority of the objects remained unidentified. For
the sources in \object{M31}, the observed luminosities range from $\sim3
\times 10^{35} \mbox{erg s}^{-1}$ to $\sim2 \times 10^{38} \mbox{erg s}^{-1}$
(at the assumed \object{M31} distance of 690 kpc used throughout this paper;
see Capaccioli et al.\ \cite{Cap89}). Also this very first survey settled the
question of whether a difference between the integrated luminosity
distribution of the globular cluster sources in M31 and the one in our own
Milky Way existed, by showing that they were in fact similar. Also, spectral
analyses of the 56 brightest sources were presented, confirming their
identifications with optical sources. Additionally, a diffuse component within
the bulge region was found, its luminosity estimated to be less than $3.2
\times 10^{38} \mbox{erg s}^{-1}$.

For 10 of the brightest sources, and for the bulge as a whole, Trinchieri et
al.\ (\cite{Tri99}) presented results from spectral analyses based on data
obtained with BeppoSAX. They confirmed that most of the sources correlate with
globular clusters and found, for all but one, Low Mass X-ray Binary (LMXB)
spectra, typical of  X-ray sources in globular clusters. Additionally, they
also confirmed the presence of two components in the spectrum of the bulge,
though they also stated that it is consistent with a superposition of many
LMXB spectra (as Irwin \& Bregman \cite{Irw99} also did using ASCA and ROSAT
data). Furthermore they extended the spectral data range up to $\sim30$ keV by
making use of the PDS instrument. Because of the lower spatial resolution of
BeppoSAX compared with ROSAT, they could not resolve the bulge into
individual sources. Garcia et al.\ (\cite{Gar00}) reported the separating of
the central source into 5 individual sources using Chandra data. They
interpreted one of these sources ($1\arcsec$ from the centre) as a possible
central black hole, although it shows an unusual (soft) spectrum. In contrast,
using ROSAT PSPC observations, Borozdin \& Priedhorsky (\cite{Bor00}) found
all the resolved X-ray sources in the core of M31 to be in accordance with LMXB
spectra, this after subtracting a soft component (thought to be thermal
emission from hot gas) derived from the area around the individual sources.

The second PSPC survey of \object{M31}, described in this paper, is an
improvement over the first in terms of its higher spatial homogeneity
across the entire disk of M31.
This has resulted here in the detection of 396
X-ray point sources, leading to a grand total of 560 individual sources
on merging the two surveys.
Both the list resulting from the here-described analysis of the second 
survey as well as the merged list of both surveys are presented
in this paper. For the majority of the
X-ray sources already identified in the first survey, the positional accuracy
has been improved. Also, the time interval of approximately one year between
the two surveys allows a time variability study to be made, and this is also
described in this paper. Spectral analysis is not presented here as it is only
suitable for the brightest sources, which were already seen in the first
survey, their spectra having been studied in S97.

\section{Observations}

\begin{figure}
   \psfig{figure=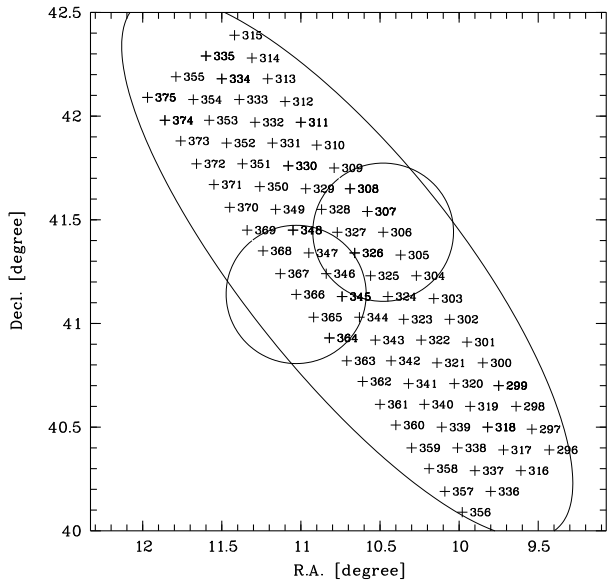,width=8.5cm,bbllx=9mm,bblly=2mm,bburx=206mm,bbury=195mm,clip=}
   \caption[]{\label{locations} Locations of the 80 on-axis pointing 
              directions of the second PSPC survey of \object{M31}. The numbers beside
              the crosses give the last three digits of the corresponding
              ROSAT observation ID.
              Two PSPC central regions with $20\arcmin$ radius are drawn
              representatively (whereas the total FOV is $\sim3$
              times larger). The $D_{\rm25}$-ellipse of \object{M31} is indicated.}
\end{figure}

The analysis in this paper is based on the second pointed \object{M31} survey
with the ROSAT PSPC, performed in July/August 1992 together with a few
follow-up observations in January 1993. This survey consists of 94
observations of 80 different pointing positions, with 2.5 ks total observation
time for each pointing.
15 follow-up observations became necessary to complete the
observation time of 14 previously interrupted observations and 1 completely
failed observation. The details of
the observations are given in Table \ref{journal}. This entire raster pointing
covered the whole disk of \object{M31} (in terms of its $D_{\rm25}$-ellipse) and
more, in 4 strips of 20 pointing directions each. Fig.\ \ref{locations} shows
the location of these 80 pointings, crosses marking the on-axis directions of 
the
telescope, and the numbers giving the last three digits of the corresponding
ROSAT observation ID. Follow-up observations are not numbered as they have the
same pointings as their corresponding main observations. For two pointings, a
$20\arcmin$ radius circle is drawn to represent the inner area of the PSPC.
The instrument's angular resolution and sensitivity are best
within this area, though the total field of view (FOV) of the PSPC is
$57\arcmin$ in radius. The $D_{\rm25}$-ellipse of \object{M31} is also drawn for
orientation. Just from this image, it can be seen that the whole of
\object{M31} is covered by the inner PSPC region, leading to an overall
approximately constant sensitivity.

The observations were performed in the normal ROSAT wobble mode which adds a
slight positional oscillation of $\sim3\arcmin$ amplitude on the nominal
pointing direction. This was done to smooth out the shadow of the PSPC support
structure and to prevent point sources being continually hidden behind the
main ribs of the structure. This wobble mode, the finely rastered array of the
pointing directions, and the fact that the whole disk of \object{M31} was
covered with the inner PSPC region explain the higher homogeneity of the
second survey compared with the first described in S97.

\begin{figure*}[t]
   \psfig{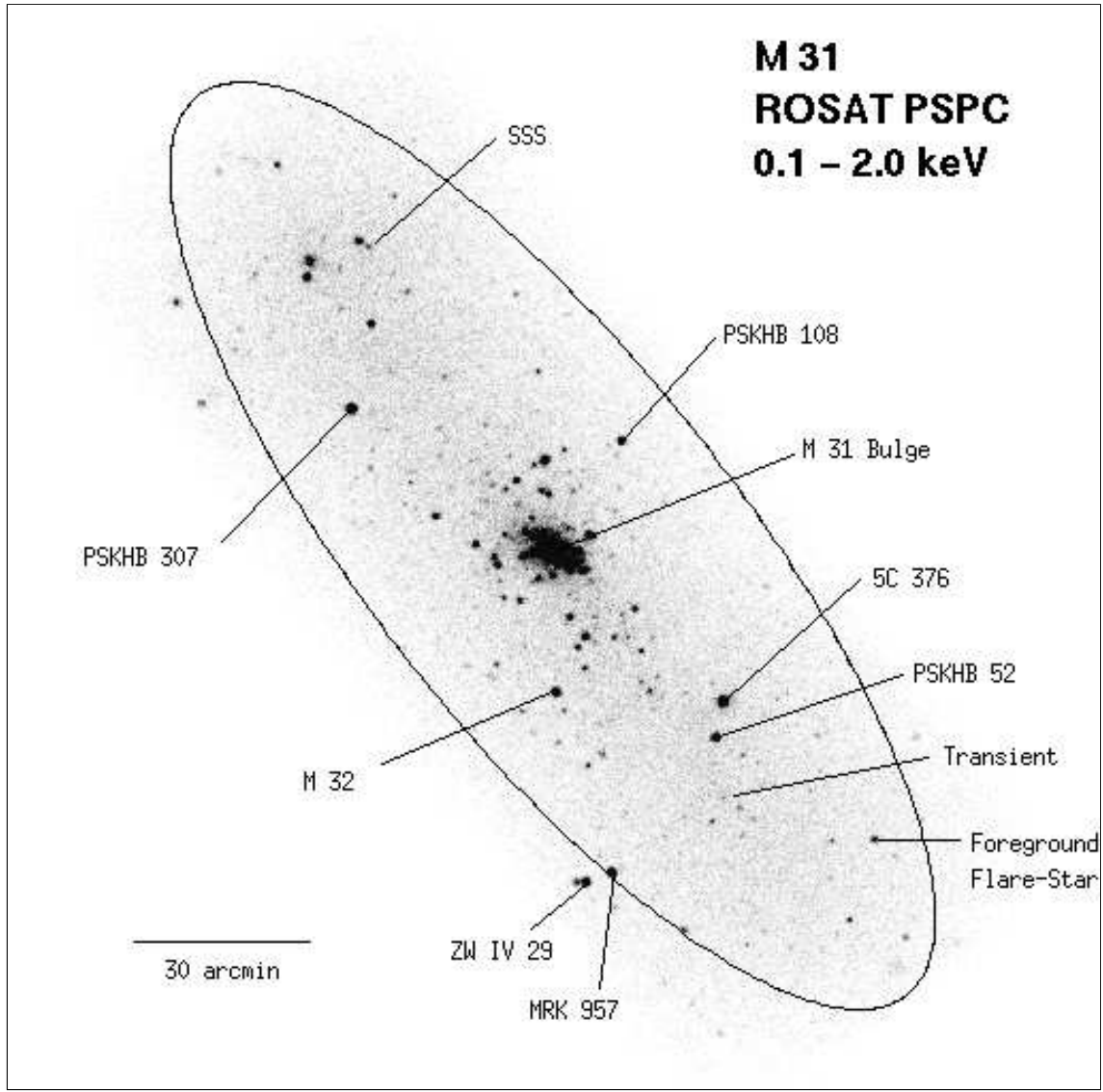}
   \caption[]{\label{rawimage} Projection of the B-band photons of the
              merged 94 pointings in the inner PSPC regions ($20\arcmin$
              radius each) with a pixel size of $21.5\arcsec \times 
              21.5\arcsec$. The $D_{\rm25}$-ellipse of \object{M31} is marked.
              For a few bright sources their identifications are given, 
              `SSS' standing for `supersoft source'. During the first
              ROSAT \object{M31} survey, the here indicated transient source 
              was as bright as \object{PSKHB 52}.}
\end{figure*}

\begin{figure*}[t]
   \psfig{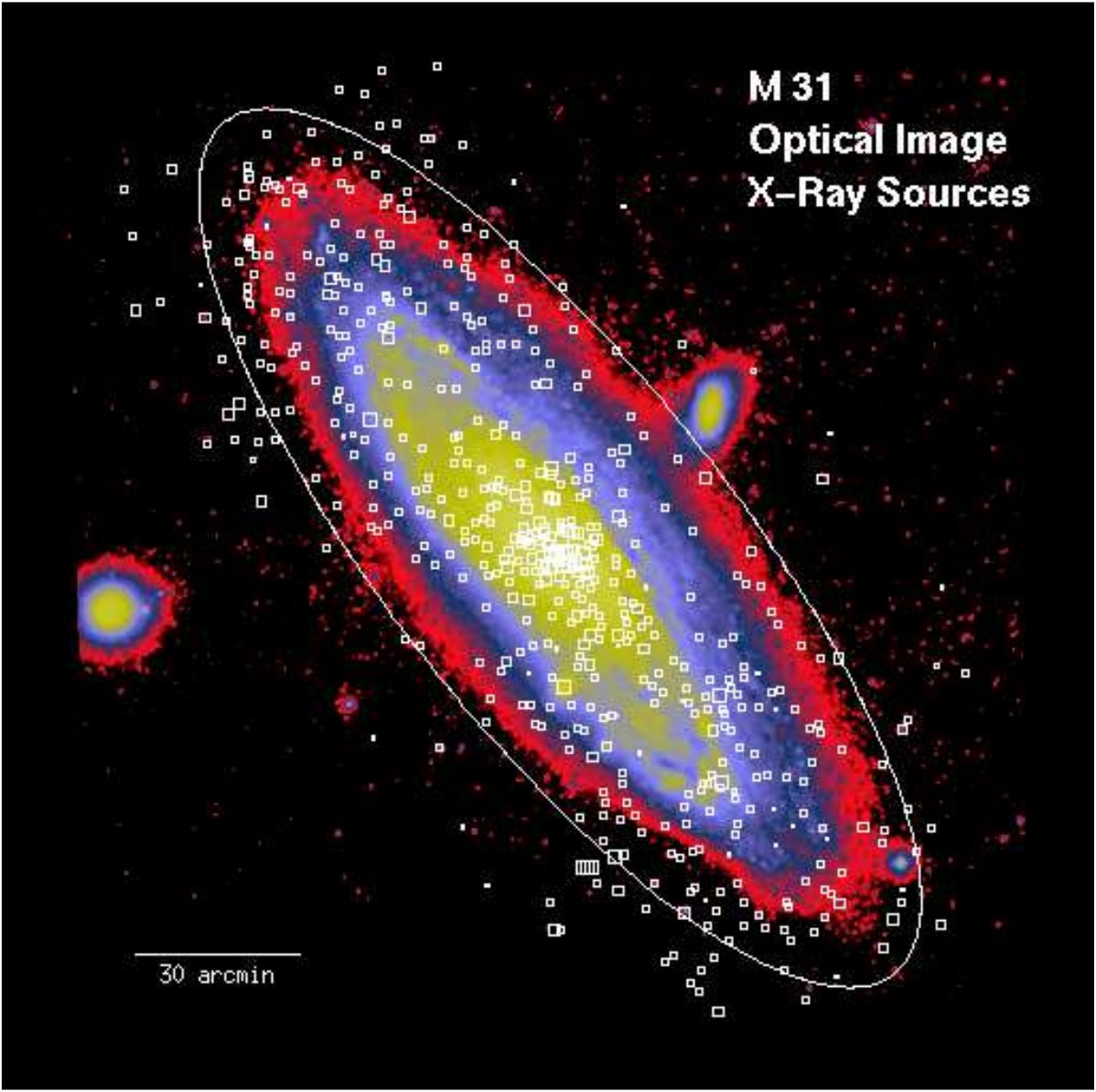}
   \caption[]{\label{optimage} False colour image of \object{M31} made from an optical
              photography with the Mount Palomar observatory. Size and orientation
              are as in Fig.\ \ref{rawimage} (also the
              $D_{\rm25}$-ellipse of \object{M31} is marked). The white
              boxes give the 560 X-ray source positions from the 
              analyses of the first and second ROSAT PSPC survey of \object{M31} as
              described in Sect.\ \ref{sourcedescript} and listed in Table
              \ref{sourcelist}. The 4 box 
              sizes indicate the logarithm of the X-ray luminosities 
              below 36, between 36 and 37, between 37 and 38, and above 38
              (from small to large). This corresponds to flux thresholds
              of $1.76 \times 10^{(-14, -13, -12)} \mbox{erg cm}^{-2} \mbox{ s}^{-1}$.}
\end{figure*}

\section{Analysis}
\label{Analysis}

For the analysis, the ROSAT energy range from 0.1 to 2.4 keV was divided into
five energy bands: a soft band (S: 0.1 - 0.4 keV), two hard bands (H1: 0.5 -
0.9 keV and H2: 0.9 - 2.0 keV), and two combined bands (hard H: 0.5 - 2.0 keV
and broad B: 0.1 - 2.0 keV). This energy band splitting was used previously in
the analysis of the first \object{M31} survey (S97), except that an upper limit
of 2.4 keV was used for the B-band. The change from 2.4 keV to 2.0 keV 
makes no significant
difference due to the drastic drop in effective area for the ROSAT telescope
+ PSPC instrumentation between 2.0 and 2.4 keV (the count rate in the 0.1-2.0
keV energy band is 2\% less than in the 0.1-2.4 keV band, when applying a
power law with photon index $\Gamma = -2.0$ and $N_{\rm H} = 9\times10^{20}
\mbox{ cm}^{-2}$ as a spectral model -- typical for \object{M31} sources).
Therefore the count rates of the two survey analyses are directly comparable.

Parts of the following analysis are based on the Extended Scientific Analysis
System (EXSAS; Zimmermann et al.\ \cite{Zim93}) developed at the
Max-Planck-Institute f\"ur extraterrestrische Physik.

\subsection{Data preparation and images}
\label{Data preparation}

All the data were inspected for contamination by solar scattered X-rays and
particle background. The first originate from Thomson and fluorescent
scattering of solar X-ray photons with atoms and molecules in the upper
atmosphere along the line of sight. For the ROSAT orbit, these are mainly
oxygen, nitrogen, argon, helium, and hydrogen (Jacchia \cite{Jac72}). For the
integral solar scatter, the illuminated column density of the atomic oxygen can be used
because of the well known fixed ratio of scatter contribution of the other
components, as discussed in detail by Snowden \& Freyberg (\cite{Sno93}).
Therefore, for each pointing, the column density of atomic oxygen was
calculated from the orientation of the telescope and the sun position during
the whole observation. All time intervals with oxygen column densities above
$1 \times 10^{15} \mbox{ cm}^{-2}$ (see Snowden \& Freyberg \cite{Sno93} for
an explanation of this threshold) were rejected.

Snowden et al.\ (\cite{Sno92}) found a strong correlation between the Master
Veto Rate of the ROSAT onboard electronics and the residual particle background
not rejected by the veto electronics. Therefore, all time intervals with a
Master Veto Rate of more than $170 \mbox{ ct s}^{-1}$ were additionally
rejected. Applying these procedures, the rest of the scattered X-rays and
residual particle background within the screened intervals was estimated to be
less than $1\%$.

For the following analysis, the photon events of all 94 observations
representing the survey were merged into one single event file. This increased
the photon statistics and allowed us to make use of the homogeneity of the
raster survey. A slight random offset and rotation of each pointing was
corrected for by first correlating bright point sources in neighbouring
pointings detected by the Standard Analysis Software System (SASS) and delivered
with the data. For this purpose, only sources within the inner PSPC region
($20\arcmin$ radius) were used where the telescope has its highest
spatial resolution. The final source position was calculated as the weighted
mean position from the individual source positions in each contributing
pointing, with the signal to noise ratio as the weighting factor. In a last
step, each contributing pointing was shifted and rotated to fit best this mean
source position. The distribution of the shift and rotation offsets over all
94 pointings was found to be gaussian-like, with $\sigma = 5.2\arcsec$ 
in shift and $\sigma = 0.21\degr$ in rotation.
These corrected data were then ready to be merged.


Fig.\ \ref{rawimage} shows a photon image in the B-band from the merged inner
PSPC regions of the 94 pointings with a pixel size of $21.5\arcsec \times
21.5\arcsec$. Just from this image, the high homogeneity and the narrow 
(center of detector) point spread function (PSF; Hasinger et al.\ \cite{Has92})
of the second ROSAT \object{M31} survey across the whole galaxy
(indicated by the $D_{\rm25}$-ellipse) can be seen, especially when compared to
the image of the first survey (Fig.\ 2 in S97). Some bright identified sources
are also indicated in Fig.\ \ref{rawimage}. Most of them are not members of
the \object{M31} system. The bulge region is severely crowded by point sources
and confused by an additional diffuse component.

Fig.\ \ref{optimage} shows an optical image (taken from the Mount Palomar
Sky Survey) of \object{M31} in false colour representation. Size and 
orientation are as in Fig.\ \ref{rawimage} and the
$D_{\rm25}$-ellipse of \object{M31} is also marked. The white boxes
mark the 560 X-ray source positions from the analyses of both
ROSAT PSPC surveys of M31 as described in Sect.\ \ref{sourcedescript} 
and listed in Table \ref{sourcelist}. The 
4 box sizes indicate the logarithm of the X-ray luminosities 
below 36, between 36 and 37, between 37 and 38, and above 38
(from small to large). This corresponds to flux thresholds
of $1.76 \times 10^{(-14, -13, -12)} \mbox{erg cm}^{-2} \mbox{ s}^{-1}$.
For flux calculations, a spectral model of a power law with
photon index $\Gamma = -2.0$ and $N_{\rm H} = 9 \times 10^{20} \mbox{ cm}^{-2}$
has been used which holds for \object{M31}-sources but not for foreground 
or background objects. A distance of 690 kpc for \object{M31} is asumed
for the resulting luminosity values.

\subsection{Source detection}

To make use of the high homogeneity of the second PSPC \object{M31} survey,
the source detection was performed on the merged data of the inner PSPC
regions of all 94 observations. This guaranteed the best results for the
determined source positions and covered approximately the whole $D_{\rm25}$-area
of \object{M31}. For detections of sources outside this region, the following
source detection procedure was repeated using the merged data of the total
FOV. The source detection technique used is similar to the one previously used
for the analysis of the first survey and described in detail in S97. Hence,
only a brief description will be given here, with emphasis on the differences
employed.

The computations can be divided into three steps: a local, a map, and a
maximum likelihood detection algorithm. For the local detect algorithm, the
merged photon event tables were split into a northern, middle and southern
part and for each part, images were created with a pixel size of $15\arcsec
\times 15\arcsec$ for each of the five energy bands. This led to $3 \times 5 =
15$ images for the three regions and the five energy bands. With a sliding
window technique ($3 \times 3$ pixel box), the images were searched for a
significant count excess within the box compared with the surroundings. Only
source candidates with a likelihood of existence $\ge 8$ were listed, where
the likelihood L = -ln(P), P being the probability that the measured number of
photons in the box originate from Poissonian background fluctuations.

In the following map detect algorithm, the same procedure was applied to the
15 images, but this time the photon number within the box was compared with
the number of photons within a box of equivalent area and position in a
background map. These background maps were computed from the photon images by
punching out holes at the source positions determined by the local detect
algorithm, and applying smoothing procedures before and afterwards as
described in S97. The radius of the holes was set to twice the FWHM of the PSF
computed for a $20\arcmin$ off-axis angle and for the lowest energy value
within the considered energy band (a $40\arcmin$ off-axis angle was used for
the merged total FOV data). This resulted in a second list of source
candidates (also with L $\ge 8$).

For the third step, the local and map source candidate lists were merged into
one list (separately for each of the five energy bands) and used as input for
a maximum likelihood detection procedure (Cruddace et al.\ \cite{Cru88}). Here
only sources with a likelihood L $\ge 10$ were accepted and the background
maps described above were used. All resulting lists were merged into one final
list such that sources separated by less than $2\sigma$ of the PSF (referring
to the lowest energy value within the considered energy band) were substituted
by one single source, its position set to the position of the original source
with the highest likelihood. This list was used as input for a repeated
maximum likelihood process to compute upper limits in the energy bands where a
source was below our detection threshold (but above in any of the other energy
bands).

\subsection{The catalogue of detected X-ray sources}
\label{sourcedescript}

The source detection yielded the 396 sources listed in Table \ref{sourcelist92},
which has the same structure as the first survey source list 
given in Table 5 of S97. Column 1 gives the source number,
columns 2 - 7 list the centroid position (epoch J2000) after
correction for a systematic offset (see below) and column 8
shows the $1\sigma$ uncertainty of the source position in arcseconds.  The
calculation of this positional uncertainty is based on the maximum likelihood
algorithm and incorporates the effects of statistical errors depending on the
number of source counts, together with the blur radius of the PSF at the
off-axis angle and the mean photon energy of the source. We also set a minimum
threshold of $5\arcsec$ to account for a systematic positional error.  The
parameter in column 9 represents a classification of the quality of the
detection and is differently defined than for the first survey due to the
different homogeneity and sensitivity of the second survey:
class `1' indicates
sources detected in the inner PSPC region ($20\arcmin$ radius) and class `4'
sources outside this region. Column 10 in Table \ref{sourcelist92} gives the
highest likelihood of existence found in any of the five energy bands computed
with the maximum likelihood method. Finally, columns 11 to 15 list the count
rates with their $1\sigma$ errors (in counts per kilosecond) within the five
energy bands (B, S, H, H1, and H2; see beginning of Sect.\ \ref{Analysis}).
The listed count rate errors are only statistical errors, whereas the
systematical errors are expected to be less than $\pm15\%$. Because some faint
sources were not detected in all energy bands (i.e., these sources had a
likelihood below the threshold value of 10 in one or more energy bands), we
present $1\sigma$ upper limits to their count rates. The upper limits are
computed from the $1\sigma$ fluctuations (Poissonian statistics) of the
background counts at the source position and are indicated by a preceding
`$<$` symbol.

The 396 X-ray sources found in the second PSPC survey underwent a correlation
with a positionally accurate (optical) reference catalogue to determine a
systematic offset in source position. This was done in the same manner as for
the sources in the first survey, and is described in detail in S97. Here, for
reference, we also used the optical globular cluster catalogue of Magnier et
al. (\cite{Mag94a}; Table 2) which revealed a slight systematic offset in our
source positions of $\Delta R.A. = 5.8\arcsec$ and $\Delta Dec. = 1.2\arcsec$.
Table \ref{sourcelist92} lists the offset-corrected source positions.

The fact that the total number of detected sources in the second PSPC survey is
identical with the total number of detected sources in the first PSPC survey
(S97) is purely accidental. The source lists are different and contain only
239 common sources. The detection of common sources in the two 
surveys is due to the fact that approximately the same region of sky was
observed over (in some areas) similar integrated exposure time. 
The differences in the
source lists are mainly due to different sensitivity characteristics: the
first survey has its highest sensitivity along a line following the main axis
of the \object{M31} ellipse, whereas the second survey has an approximately
constant and high sensitivity across the whole galaxy. Therefore, the detected
sources are concentrated within different regions of each survey.
Additionally, the slight differences in the source detection procedures and
statistical fluctuations cause some departures close to the detection
threshold. 

Merging of the two survey lists (see Sect.\ \ref{corSurveyI} for details)
yielded a final catalogue containing 560 PSPC
detected X-ray sources in the field of \object{M31}. This is presented in Table
\ref{sourcelist}, which has a similar structure to Table \ref{sourcelist92}
described above. The only differences are that column 1
gives the RXJ-number of the source and column 2 lists the source number of the
first survey (as listed in Table 5 of S97) if a correlation was found. Here
four possible cases are indicated; (i) number followed by `+': source was
found in both surveys and the listed data are from the first survey, (ii)
number followed by `-': source was found in both surveys and listed data are
from the second survey, (iii) number without any additions: source was found
only in the first survey, the listed data being from there, and (iv) no number
at all: source was found only in the second survey, the listed data being from
there. For the criteria
of which data are listed in cases of correlation see Sect.\ \ref{corSurveyI}.
The following columns 3 - 16 are identical with columns 2 - 15 of Table \ref{sourcelist92},
and have been described above. For sources found in the first survey,
the classification parameter listed in column 10 is as follows: 
class `1' identifies sources detected in the
central region of the PSPC with off-axis angles $\le 20\arcmin$, class `2'
defines locations of sources found between $20\arcmin$ and $40\arcmin$, and
class `3' contains sources with off-axis angles $> 40\arcmin$. As mentioned in
Sect. 2 of S97, the source position was derived from the pointing in which it
appears at the lowest off-axis angle, i.e., the best class (though not under a PSPC
rib). For sources in class `2' and especially class `3', any upper limit in
count rate listed in columns 11 - 15 can even be an underestimation 
due to the wider PSF and the therefore
higher possibility of rib influencies.
For sources found in the second survey, the listed classification
parameter for the quality of detection is defined as described above: 
class `1' for
sources detected in the inner PSPC region ($20\arcmin$ radius) and class `4'
for sources outside this region.

The caveats for the first survey source catalogue mentioned in S97 are still
valid where these sources are not substituted by second survey detections.

\section{Comparisons with other source catalogues}
\label {correlations}

For all correlations with other catalogues described in this
section, the final source list of Table \ref{sourcelist} 
was used. Table \ref{corover} summarises the
results of the correlation analysis for different catalogues and these are
discussed in more detail in the following subsections. From the description
(in S97) of the correlation process itself, we simply summarise here that it
yields not only the total number of correlating sources ($N_{\rm total}$) but also
the amount of expected accidental correlations ($N_{\rm acc.}$) within a $1\sigma$
confidence level.

\begin{table}
\small
\caption[]{\label{corover} Summary of the correlation analysis. $N_{\rm total}$
                          gives the number of all possible correlations within a distance
                          of $2\sigma$ of the combined positional error, $N_{\rm acc.}$ gives
                          the number of statistically expected accidental correlations,
                          and $N_{\rm fin.}$ gives the final
                          accepted correlations. For a detailed explanation
                          see Sect. \ref{corall}.}
\begin{flushleft}
\begin{tabular}{llrrr}
\hline
Type & Databases & $N_{\rm total}$ & $N_{\rm acc.}$ & $N_{\rm fin.}$\\
\hline
X-ray         & {\it Einstein} (TF) & 82 & 12.7 & 69\\
GC            & BA87, BA93, MA94a   & 43 & 11.6 & 33\\
Extragalactic & NED                 & 10 &  0.6 & 10\\
Foreground    & MA92, SIMBAD        & 72 & 40.4 & 55\\
SNR           & DO80, BW93, MA95    & 22 &  4.1 & 16\\
Novae         & SA91, SA92          &  0 &  0.8 & --\\
\hline
\end{tabular}
\end{flushleft}
\end{table}

\subsection{Comparisons with previous X-ray source catalogues}

\subsubsection{Comparison with the first ROSAT survey of \object{M31}}
\label{corSurveyI}

The source list of the second \object{M31} survey was merged with that of the
first to obtain the final ROSAT PSPC X-ray source list of \object{M31} (Table
\ref{sourcelist}). For this purpose the above-mentioned correlation process
was applied to both lists to identify common sources. The `radius of
acceptance' ($r_{\rm a}$) -- the important correlation parameter -- was iteratively
determined as follows: the correlation procedure was repeatly carried out,
with $r_{\rm a}$ increasing successively from $r_{\rm a} = \sigma_{\rm comb.}$ (here,
$\sigma_{\rm comb.} = \sqrt{\sigma_{\rm 1}^2 + \sigma_{\rm 2}^2}$, the combined positional
error of the correlating sources, where the single positional error of each
source is given by the maximum likelihood detect algorithm). This iterative
process was stopped just before the occurrence of multi-identifications (one
source in one catalogue correlating with more than one source in the other
catalogue) for sources with likelihood $\ge 20$ (to exclude sources near the
detection threshold), and outside confused regions. In this way, most
potential common sources have been uncovered, without risking having to accept
multi-identifications for bright isolated point sources.

This process yielded 239 correlations (with $r_{\rm a} = 4 \sigma_{\rm comb.}$,
corresponding to a 99.99\% probability of all real identifications having been
found) with an expected number of 23 chance coincidences ($1\sigma$-value). A
few multi-correlations were accepted (see below) because they either occur
within confused regions or have likelihoods $< 20$ or at least one of the
correlating sources is covered by the PSPC rib structure.

For sources correlated within the two surveys, the one with the better quality
of detection (i.e. with the lowest classification parameter listed in column
10 of Table \ref{sourcelist}, class `4' of the second survey being considered
the same as classes `2' and `3' of the first) was taken and the other was
rejected as being identical with the first. In cases where the correlating
sources were of the same class, the one with the higher likelihood was taken
and, if the likelihood was also the same, the source from the second survey
was taken because of its better positional accuracy\footnote{As a concequence
of this data quality based decision variable sources are preferentially listed 
in their luminous state which may have different spectral characteristics
compared with their less luminous state. Therefore sources detected as
variable (see Tables \ref{varEinsteinList} and \ref{varSurveysList}) are marked 
with a `$\sim$`-sign in front of their RXJ-number.}. The same procedure was
applied to the few multi-correlations to clarify their situation.
With this only one multi-correlation remained: source \#379 of the first
survey (see Table 5 in S97) correlates with sources \#380 and \#384
of the second survey (see Table \ref{sourcelist92} in this paper). Applying
the rules mentioned above for both correlations we would have to accept 
source \#379 from the first survey and would have to reject both sources 
from the second survey. Here we decided only to accept the correlation
with the least distance as a true identification and left source \#384
as a new one.

Conversely, 158 sources from the first survey and 163 from the second do not
correlate with any other source (and we extend to 164 for the second survey 
due to the reasons mentioned above). To consider all these sources as transients
would ignore the different spatial sensitivity distributions and different sky
coverage of the two surveys. Therefore, a more explicit investigation of
transients is presented in Sect.\ \ref{Variability}.

\subsubsection{Comparison with the {\it Einstein} catalogue}
\label{corein}

The 560 X-ray sources in the merged source list of the two ROSAT PSPC surveys
exceeds the number of X-ray sources detected with the {\it Einstein}
observatory in this region of sky by a factor of more than 5. On the one hand,
it is the result of the $\sim10$ times higher sensitivity of ROSAT and the
larger exposure of the disk region in the second ROSAT survey.
On the other hand, both ROSAT surveys covered a more complete and therefore
larger portion of the \object{M31} field than the {\it Einstein}
observations did. The number of sources detected with the ROSAT PSPC in the
\object{M31} bulge region (within 1 kpc from the centre) increased from 22
in the first survey to 31 using the data from both surveys. The fact that this
is still less than the 48 sources found with the {\it Einstein} observatory in
this region, as listed by Trinchieri \& Fabbiano (\cite{Tri91}, hereafter TF), 
is due to the large fraction of sources in TF's list which were detected with
the higher spatial resolution {\it Einstein} HRI. Primini et al.\
(\cite{Pri93}) reported 45 sources found with the ROSAT HRI within the bulge
region of \object{M31} and Immler (\cite{Imm00}), again using the ROSAT HRI
observations, counted 63 sources within a $5\arcmin$ circle around the centre.

As already described in S97, the list of {\it Einstein} X-ray sources in the
field of \object{M31} reported by TF contains 108 sources, with 81 sources
taken from the {\it Einstein} HRI data with an assumed positional error of
3\arcsec \ (reported by Crampton et al. \cite{Cra84}), and 27 sources based on
{\it Einstein} IPC data with a 45\arcsec \ positional error. Applying the
above mentioned correlation procedure to the 560 ROSAT sources and the 108
{\it Einstein} sources reported by TF yields $N_{\rm total} = 82$ correlations
with a probable contamination of $N_{\rm acc.} = 12.7$ chance coincidences, here 
accepting
a source separation of up to twice the combined positional error ($2\sigma$).
12 ROSAT sources each correlated with several {\it Einstein} sources,
due mainly to the large positional error of the {\it Einstein} IPC. To
clarify their situation, only the correlation with the smallest separation
(between the correlating counterparts) was taken into account. This reduced
the number of finally accepted correlations to 69, which is in good agreement
with the number of statistically expected true correlations (i.e. $N_{\rm total} -
N_{\rm acc.} = 69.3$).

All 69 identifications are listed in Table \ref{ID-Einstein}. Column 1 gives
the ROSAT RXJ-number (ref. Table \ref{sourcelist}), column 2 gives the fluxes
and $1\sigma$ errors of the ROSAT sources using the spectral model of TF
(thermal bremsstrahlung with $kT = 5 \mbox{ keV}$ and $N_{\rm H} = 7 \times
10^{20} \mbox{ cm}^{-2}$ in the 0.2-4.0 keV energy band), column 3 lists the
{\it Einstein} source numbers (ref. Table 2A of TF), column 4 the fluxes and
$1\sigma$ errors given by TF, and columns 5 and 6 the distances between the
ROSAT source positions and the {\it Einstein} source positions in arcseconds
and in units of their combined positional errors ($\sigma$) respectively.  The
last column shows the ratio between the fluxes obtained with ROSAT and {\it
Einstein} and can be considered as a long term variability check between the
epochs of the two observations. More detailed investigations into long time
variabilities are described in Sect.\ \ref{Variability}.

Comparing this correlation list to the one using only ROSAT sources found in
the first survey as published in S97 (Table 6), a few remarks should be made.
Using only the data of the first survey we had to manually extend the
correlation list by one entry (ROSAT source \#67 correlating with {\it
Einstein} source \#3) as mentioned in S97. This had been necessary because of
the poorly--known PSF and the therefore uncertain positioning at the source
position. The second PSPC survey now gave us the
opportunity to determine much more precisely the position of this source 
(RXJ0040.2+4050), turning out in fact to be only $3.6\arcsec$ away from
the position of the {\it Einstein} HRI source \#3. Therefore no manual
extension of the correlation list had to be made in this paper.

The listed flux ratio ($F_{\rm R}/F_{\rm E}$) between ROSAT and {\it Einstein} which can
be used as a long term variability indicator should be inspected carefully for
sources in the bulge region (marked with a $\star$ preceding the ROSAT source
number). Because of the heavy confusion in this region, the flux determination
of these sources is very uncertain.

With the help of the second PSPC survey, some positions
of X-ray sources already found in the first survey could be improved.
Therefore, the 69 identifications listed in Table \ref{ID-Einstein} show
a very good positional agreement between the PSPC source positions and
the ones listed by TF, which were largely obtained with the {\it Einstein} HRI.
In fact, the mean source separation of the 43 ROSAT PSPC--detected sources
correlating with sources also found with the {\it Einstein} HRI is
$5.9\arcsec\pm3.2\arcsec$.

Excluding the heavily confused bulge region and the sources therein, we found
a good ROSAT confirmation (90\%) of the sources detected with the {\it
Einstein} observatory as, out of the 60 of the 108 {\it Einstein} sources
outside the bulge region, 54 could be confirmed by ROSAT. For the 6 {\it
Einstein}-only detected sources, we give ROSAT flux upper limits and discuss
their transient nature in Sect.\ \ref{VarEinstein}. Over and above this, 491
new sources have been found with ROSAT which were not detected with {\it
Einstein}.

\subsection{Correlations with optical and radio sources}
\label{corall}

To identify and classify individual sources, the merged ROSAT source list of
both surveys (Table \ref{sourcelist}) was correlated with the same catalogues
previously used for the sources of the first survey in S97. For completeness
and to simplify the discussions, we summarise the public data bases and
catalogues used as follows:
\begin{itemize}
        \item {\bf globular clusters:} the two lists of Battistini et al.
                        (\cite{Bat87}, \cite{Bat93}; hereafter BA87, BA93) and the lists of
                        Magnier et al. (\cite{Mag94a}; Table 2; hereafter MA94a),
 
        \item {\bf extragalactic objects:} the NASA Extragalactic Database
                        (version date: 30. Dec. 1992; hereafter NED),

        \item {\bf foreground stars:} the catalogue of stellar
                        photometry described by Magnier et. al. (\cite{Mag92}), 
                        hereafter MA92, and
                        Haiman et al. (\cite{Hai94}) and the
                        SIMBAD catalogue (Centre de Donn\'ees astronomiques de Strasbourg;
                        version date: Dec. 1989; hereafter SIMBAD),

        \item {\bf supernova remnants:} the lists of d'Odorico et al.
                        (\cite{Dod80}; hereafter DO80), Braun \& Walterbos
                        (\cite{Bra93}; hereafter BW93), and Magnier et al.
                        (\cite{Mag95}; hereafter MA95).

        \item {\bf novae:} the two lists of Sharov \& Alksnis
                        (\cite{Sha91}, \cite{Sha92}; hereafter SA91, SA92).
\end{itemize}
Information regarding the characteristics of these catalogues, especially the
individual positional errors used in the correlation processes, can be found in
S97. We adopted them except for the SNR catalogues: D'Odorico et al.\ (\cite{Dod80})
report general position errors of $8\arcsec$ in declination and $15\arcsec$
in right ascension. In S97 we assumed a mean position error of $12\arcsec$
whereas in this paper we decided to use a geometric mean of $17\arcsec$.
For the SNR list of Braun \& Walterbos (\cite{Bra93}) and also 
for the list of Magnier et al.\ (\cite{Mag95}) we used $5\arcsec$ as
a systematic position error for our correlations.

Table \ref{ID-Tabelle} shows the result of the correlations. The columns are
defined as follows. Column 1 gives the ROSAT RXJ source number (ref. Table
\ref{sourcelist}). Column 2 lists the object class, of which four exist:
`Star' for galactic foreground stars followed in brackets by their type if
available, `EO' for extragalactic objects, mainly background galaxies, `GC'
for sources belonging to globular clusters, and `SNR' for supernova remnants.
Column 3 lists the identification, using the abbreviations of the correlated
catalogues as defined above. The number following in brackets gives the
name/entry number of the object as listed in the relevant catalogue (for
details see the remarks to the individual catalogues below). Finally, columns
4 \& 5 give the distance between the ROSAT source position and the correlated
object in the catalogue, both in arcseconds and in $\sigma$ units. For the
distance expressed in sigma, the combined positional error of the ROSAT source
and the correlated catalogue source was used.

Concerning this list, the following should be noted. If one ROSAT source
correlates with more than one catalogue source of the same catalogue, only the
correlation with the smallest positional separation is listed. If the
correlating catalogue sources belong to different catalogues of the same
source class then all correlations are listed, separated by commas. In a few
cases, multi-correlations between one ROSAT source and catalogue sources of
different source classes were found. Here, spectral considerations clarified
the situation, especially for distinguishing between foreground stars and
globular clusters. 
Rejections of a good spatial correlation in place of a poorer spatial 
correlation only took place when the more distant counterpart was spectrally 
consistent with the ROSAT source and the closer counterpart very inconsistent.

In contrast, no rejection was performed in cases of perfect positional
single--correlations, even of moderate coincident spectral characteristics.
Additionally, we did not accept identifications with supernova remnants for
ROSAT sources with a hardness ratio $HR_{\rm 1} + \sigma_{\rm HR_1} \le -0.80$, because
we consider these sources as supersoft sources (see S97 and Greiner et al.\
\cite{Gre96})\footnote{Kahabka (\cite{Kah99}) used not only $HR_{\rm 1}$ but also
$HR_{\rm 2}$ and $L_{\rm X}$ (with reference to the local $N_{\rm H}$-value at the
source position) and their respective ratios to discriminate supersoft
sources. Using these criteria and the source list of S97, he came up with an
additional 26 new supersoft source candidates, 4 of them correlated with
foreground stars in S97 and 4 with supernova remnants. Excluding these 8
sources, 18 additional supersoft source candidates in M31 remain.}. 
The hardness ratio $HR_{\rm 1}$ is defined as $HR_{\rm 1} = (H - S) / (H + S)$, where $S$
and $H$ stand for the source counts in the relevant energy bands calculated
with the maximum likelihood algorithm (and listed in Table \ref{sourcelist}).
With these criteria, 114 identifications with optical and radio sources were
found, corresponding to an identification quota of 20.4\%. Some quantitative
comments on the various object classes are as follows:\newline
{\bf Foreground Stars (Star):} Among the $N_{\rm total} = 72$ correlations within
the $2\sigma$ error level, 17 had to be rejected due to the above criteria,
leading to $N_{\rm fin.} = 55$ finally accepted identifications. The high density
of foreground stars within the HA94 catalogue yields a relatively high number
of possible chance coincidences, $N_{\rm acc.} = 40.3$. The resulting statistically
expected number of true identifications is $N_{\rm i} = 31.8\pm6.3$, which is too
low when compared with the finally accepted 55 identifications. As already
discussed in S97, from the {\it Einstein} and ROSAT medium and deep surveys we
know the foreground source luminosity function, and this can be used to derive
an upper limit of 54 expected foreground sources within the region covered by
the HA94--catalogue. This value is in good agreement with our finally accepted
number of identifications. \newline
{\bf Background Galaxies (EO):} None of the $N_{\rm total} = 10$ found
correlations had to be rejected due to the above criteria. The remaining
number of $N_{\rm fin.} = 10$ finally accepted identifications is in good
agreement with the statistically expected number of $N_{\rm i} = 9.4\pm0.8$. The
dwarf galaxy \object{M32} can be found among the identifications, correlating
with ROSAT source RXJ0042.6+4052.\newline
{\bf Globular Clusters (GC):} Within the $2\sigma$ error level we found
$N_{\rm total} = 43$ correlations (with $N_{\rm acc.} = 11.6$ chance coincidences), from which
10 had to be rejected due to the above criteria. The remaining 33 finally
accepted identifications are in good agreement with the statistically expected
number of $N_{\rm i} = 31.4\pm3.4$ true identifications. Among the 10 rejected
correlations, 2 accounted for double--correlations with globular clusters,
while the remaining 8 were rejected on spectral grounds, showing soft spectral
characteristics incompatible with the known relatively hard spectra of X-ray
sources belonging to globular clusters. \newline
{\bf Supernova Remnants (SNR):} Among the $N_{\rm total} = 22$ correlations within
the $2\sigma$ error level, 6 had to be rejected due to the above criteria,
leading to $N_{\rm fin.} = 16$ finally accepted identifications.
This is in good agreement with the statistically expected number of $N_{\rm i} =
17.9\pm2.0$ true identifications. One important comment concerning
the SNR-correlations listed in S97: Due to a misuse of the SNR list
of Magnier et al.\ (abbreviated as MA94b in S97) a few SNR miss-correlations
are listed in S97. This is repaired in this paper.\newline
{\bf Novae:} The extension of the ROSAT source catalogue of the first
\object{M31} survey (S97) with the sources found in the second survey does not
uncover a single correlation with one of the known novae in \object{M31}.

\section{Time variability}
\label{Variability}

The two ROSAT PSPC surveys of \object{M31}, separated by $\sim1$ year, and the
{\it Einstein} survey from $\sim11$ years before the first ROSAT survey can be
used to search for long term variability within the sources. We treat this
here in two different subsections. Readers who wish to investigate long term
variabilities or the search for transients should refer to both subsections
(\ref{VarEinstein} and \ref{varSurveys}), and are strongly recommended to read
Sect.\ 4.1 in S97.

Concerning any two catalogues 1 and 2 which refer to the same sources, we used
for a quantitative study of possible long term variabilities a linear
significance parameter following Primini et al.\ (\cite{Pri93}), which is
defined as:
   \begin{equation} \label{S} S(F_{\rm 1} - F_{\rm 2}) = \frac{|F_{\rm 1} - F_{\rm 2}|}{\sqrt{\sigma^2_{\rm F_1} + \sigma^2_{\rm F_2}}} \, , \end{equation}
where $F_{\rm 1}$ and $F_{\rm 2}$ represent the X-ray flux in the first and second source
catalogues and $\sigma_{\rm F_1}$ and $\sigma_{\rm F_2}$ give the corresponding flux
errors.
This definition is useful in that, in cases where an inappropriate spectral 
model has been used to compute the two fluxes, any systematic errors are 
disregarded. 
We state time variability only for sources with $S \ge 3\sigma$.
Additionally, sources within the bulge and other confused or `handicapped'
regions (e.g. beneath the ribs of the PSPC support structure) were excluded on
cautionary grounds.

\subsection{Comparison with the {\it Einstein} sources}
\label{VarEinstein}

\begin{table}
\caption{\label{varEinsteinList} List of X-ray sources showing flux variability between the {\it Einstein}
                          observation and the ROSAT observations. $F_{\rm R}$ gives the ROSAT
                          source flux using the {\it Einstein} spectral model of TF (thermal bremsstrahlung with 
                          $kT = 5 \mbox{ keV}$ and $N_{\rm H} = 7 \times 10^{20} \mbox{ cm}^{-2}$ in the 0.2-4.0 keV
                          energy band) and
                          $F_{\rm E}$ gives the {\it Einstein} source flux of the correlated
                          {\it Einstein} source. Column ``$S$'' lists the
                          significance of the variability as described in the text.
                          A ``T'' in this column indicates bright transients or possible
                          faint transients when enclosed in brackets
                          (see Sect. \ref{VarEinstein} for a detailed explanation).}
\begin{flushleft}
\scriptsize       
\begin{tabular}{crrrrr} 
\hline
ROSAT  & \multicolumn{1}{c}{$F_{\rm R} \, (\times 10^{13})$} &
   {\it Ein.} & \multicolumn{1}{c}{$F_{\rm E} \, (\times 10^{13})$} &
   \multicolumn{1}{c}{$F_{\rm R} / F_{\rm E}$} & $S$ \\
RXJ-No. & \multicolumn{1}{c}{(cgs)} & \multicolumn{1}{c}{No.} & \multicolumn{1}{c}{(cgs)} & \\ \hline
			0040.2+4034  & $  19.87 \pm  0.29 $ &     & $           <10.00 $ &                $          >2.00 $ &    T  \\ 
			0041.7+4134  & $  14.38 \pm  0.49 $ &   9 & $   8.72 \pm  1.08 $ &                $  1.65 \pm 0.21 $ &  3.66 \\
			0041.8+4021  & $  24.25 \pm  0.76 $ &  11 & $  15.54 \pm  0.88 $ &                $  1.56 \pm 0.10 $ &  5.69 \\
			0042.2+4019  & $  40.38 \pm  1.23 $ &  15 & $  48.83 \pm  1.61 $ &                $  0.83 \pm 0.04 $ &  3.89 \\
			0042.2+4101  & $   9.53 \pm  0.32 $ &  16 & $   3.88 \pm  0.75 $ &                $  2.46 \pm 0.48 $ &  4.48 \\
			0042.2+4112  & $   9.04 \pm  0.25 $ &  19 & $   4.26 \pm  0.54 $ &                $  2.12 \pm 0.28 $ &  4.45 \\
			0042.2+4118  & $   9.71 \pm  0.32 $ &  14 & $   3.23 \pm  0.51 $ &                $  3.01 \pm 0.49 $ &  6.09 \\
			0042.6+4052  & $  32.45 \pm  0.53 $ &  51 & $   9.16 \pm  1.01 $ &                $  3.54 \pm 0.40 $ & 15.33 \\
			0042.8+4131  & $  18.62 \pm  0.45 $ &  67 & $  11.95 \pm  1.10 $ &                $  1.56 \pm 0.15 $ &  4.30 \\
			0043.1+4118  & $   7.54 \pm  0.24 $ &  82 & $   2.03 \pm  0.31 $ &                $  3.72 \pm 0.58 $ &  6.69 \\
			0043.3+4117  & $   4.44 \pm  0.25 $ &  88 & $   1.31 \pm  0.34 $ &                $  3.39 \pm 0.89 $ &  3.66 \\
			0046.4+4201  & $  10.08 \pm  0.31 $ & 105 & $   5.52 \pm  0.89 $ &                $  1.83 \pm 0.30 $ &  3.33 \\
		 &SI:  $< 0.35$ &  12 & $ 2.56 \pm 0.50$ & $       < 0.14$ &  (T)\\
		 &SI:  $< 0.44$ &  40 & $ 1.59 \pm 0.62$ & $       < 0.28$ &  (T)\\
		 &SII: $< 2.01$ &  75 & $ 4.02 \pm 0.56$ & $       < 0.50$ &  (T)\\
		 &SI:  $< 0.40$ &  84 & $ 1.99 \pm 0.49$ & $       < 0.82$ &  (T)\\
		 &SI:  $< 0.35$ &  96 & $ 3.50 \pm 0.94$ & $       < 0.10$ &  (T)\\
		 &SI:  $< 0.39$ & 106 & $ 0.71 \pm 0.22$ & $       < 0.55$ &  (T)\\
\hline
\end{tabular}
\end{flushleft}
\end{table}

As already described in Sect.\ \ref{corein}, we compared the complete ROSAT
PSPC source list of \object{M31} (Table \ref{sourcelist}) with the {\it
Einstein} source list published by TF. The results are listed in Table
\ref{ID-Einstein}, where, besides the fluxes (using the spectral model of TF),
the flux ratios are also given. Here, we extend these calculations by the
significance parameter given in formula (\ref{S}), where catalogue 1 is set to
the ROSAT source list and catalogue 2 is set to the {\it Einstein} source
list. Applying the criteria mentioned above to accept sources only with $S \ge
3\sigma$ and outside confused regions, we come up with the remaining sources
listed in Table \ref{varEinsteinList}. Additionally, this table contains
potential transients (see below). The meanings of the columns are: columns 1
and 3 give the ROSAT source number (RXJ; see Table \ref{sourcelist}) and the
correlating {\it Einstein} source (TF's source list) respectively, columns 2
and 4 list the (unabsorbed) flux and flux error of the sources as measured
with ROSAT and {\it Einstein} respectively, the spectral model of TF 
having been 
applied (thermal bremsstrahlung with $kT = 5 \mbox{ keV}$ and $N_{\rm H} =
7 \times 10^{20} \mbox{ cm}^{-2}$ in the 0.2-4.0 keV energy band), column 5
lists the flux ratio between the ROSAT and the {\it Einstein} observations,
and column 6 gives the significance parameter as described above, or a
transient indicator `T' (see below).

\noindent
{\bf Variable sources:}\newline Table \ref{varEinsteinList} lists 11 (long
term) variable sources. From a comparison between the {\it Einstein} detected
sources reported by TF and the sources found in the {\it first} ROSAT survey
of \object{M31} we reported 15 potentially variable sources in S97. Actually,
10 of the S97-reported 15 sources vanish from the variability list, and 6 new
variable sources join the list. Among the 10 vanished sources, 6 lay within
the bulge region ({\it Einstein} sources \#33, \#58, \#68, \#76, \#79, and
\#80) and have therefore been rejected from our very stringent list (we were
not so restrictive for Table 3 of S97). For 2 sources ({\it Einstein} sources
\#70 and \#348), the fluxes of the corresponding ROSAT sources have been
substituted with the data from the second PSPC survey, which were closer to
the {\it Einstein} fluxes, and the significance of variability therefore fell
below our threshold. {\it Einstein} source \#2 now correlates with ROSAT
source RXJ0040.0+4031 (formerly ROSAT source \#55) instead of ROSAT source
RXJ0040.0+4033 (formerly ROSAT source \#57) because we obtained an improved
position from the second PSPC survey data, cancelling the prior
correlation. Finally, we deleted by hand the correlation pair of {\it
Einstein} source \#27 with ROSAT source \#172 
because it lies close to the bulge within a confused region.

Among the 6 new variable sources, 3 came into the list due to their newly
determined fluxes from the second PSPC survey data (ROSAT sources
RXJ0041.8+4021, RXJ0043.1+4118, and RXJ0046.4+4201), 2 joined the list because
of the now improved positions of the correlating ROSAT sources (RXJ0042.2+4112
and RXJ0042.2+4118) and the last one (RXJ0043.3+4117) was newly discovered
within the second survey data.

In cases where a change in determined flux (between the first and second PSPC
survey) is responsible for changes in the variable source list, one should
bear in mind that this might be due to a {\it real} flux change (variability)
of the particular source within the time gap between the two ROSAT surveys
($\sim$1 year). In assembling Table \ref{varEinsteinList}, we assumed that the
changes are due to the better flux determination within the data of the second
PSPC survey compared to the first. Readers who wish to investigate the
variable sources are therefore recommended to examine all sources in
both lists.

The two variable sources reported by Collura et al.\ (\cite{Col90}) have been
discussed already in S97. Including the second PSPC survey data has added
nothing of significance as regards these.

\noindent
{\bf Transients:}\newline
Table \ref{varEinsteinList} lists 7 possible (bright) transient sources. We
define bright transients as those sources which are detected in one catalogue,
and are bright enough to be detected in the other, but which are not seen. 
ROSAT sources with fluxes $\ge 10^{-12} \mbox{ erg cm}^{-2} \mbox{ s}^{-1}$
(applying the spectral model of TF) should have been seen during the {\it
Einstein} observations. Conversely, {\it Einstein} sources with fluxes $\ge
10^{-12}\mbox{ erg cm}^{-2} \mbox{ s}^{-1}$ should have been seen in the ROSAT
surveys.

From a comparison between the {\it Einstein} detected sources reported by TF
and the sources found in the {\it first} ROSAT survey of \object{M31}, we
reported 9 potentially transient sources in S97. In detail, we have now `lost'
5 of these transients, 3 of them because the relevant {\it Einstein} sources
(\#81, \#93, and \#100) were found to correlate with sources detected within
the second PSPC survey data, and the other two because they lay within
confused regions. On the other hand, we included 3 new transients in our list
({\it Einstein} sources \#12, \#75, and \#84) because, within the first PSPC
survey their positions were near the PSPC support structure and therefore we
formally excluded them from the list at that time. With the help of the second
PSPC survey and its more homogeneous exposure, we were able to verify their
potential transient nature. For all transients, we list in Table
\ref{varEinsteinList} a flux upper limit. In the case of the ROSAT fluxes, we
compute these limits from the known background fluxes at the source positions
making use of the most sensitive survey (indicated by SI/SII for the
first/second PSPC survey). Although 3 sources were 
partially obstructed by the
PSPC support structure within the first survey (see above), for 2 of them we
calculated their upper limits from these data because these positions still
received  more exposure within the first survey than within the second. In
these cases, we simply used the second survey and its homogeneity as a proof to
clarify their transient nature.

We list all 6 transients at the bottom of Table \ref{varEinsteinList}
as faint transients (`T' within brackets) as they have luminosities
below our bright transient threshold given above, even though their {\it Einstein}
luminosities are above the detection threshold of the ROSAT surveys.

For the one transient in Table \ref{varEinsteinList} not seen by {\it
Einstein} (ROSAT source RXJ0040.2+4034), we give our transient threshold of
$10^{-12} \mbox{ erg cm}^{-2} \mbox{ s}^{-1}$ as an upper limit because TF did
not mention the limiting flux of the individual {\it Einstein}
observations. With this value, we are surely above the sensitivity of the {\it
Einstein} observations.

\subsection{Comparison between the two ROSAT PSPC surveys}
\label{varSurveys}

In Sect.\ \ref{sourcedescript} we described the merge of the two source lists
assembled from the first and second PSPC surveys of \object{M31}. Sources
which were found in both lists have been tested for variability in flux. To
indicate a possible variability we have applied the following criteria: (1)
The source must reside outside the bulge and outside other confused regions,
(2) the significance parameter (eq.\ (\ref{S})) must hold with $S \ge 3$
($F_{\rm 1}$ and $F_{\rm 2}$ being the fluxes of the source determined from the first and
second surveys), (3) sources with an upper limit to the count rate in the
B--band in either of the two surveys have been excluded (in other words, the
count rate must have been determinable), (4) sources behind/near the PSPC
support structure within the first survey have been excluded (i.e. sources
marked with a $\dagger$-symbol in Table 5 of S97, (5) the sources have to
belong to source class `1' in both surveys, and (6) the detection likelihood
of the source has to be $\ge 20$ in both surveys. Criterion (1) prevents any
pseudo-variability occurring due to uncertain flux determinations within
confused regions, criterion (2) ensures a sufficient significance, and with
criteria (3) to (6), the influence of any systematic errors should be widely
excluded.

\begin{table}
\caption{\label{varSurveysList} List of all the potentially long term variable
         sources found via a comparison of the first and second ROSAT
         PSPC surveys of \object{M31}. Column ``$S$'' lists the
         significance of the variability as described in text.
         Additionally, possible transients are tabled,
         marked with `T' in this column.} 
\tablehead{\hline
ROSAT  & \multicolumn{1}{c}{$\rm Rate_{\rm SI}$} & \multicolumn{1}{c}{$\rm Rate_{\rm SII}$} & 
			\multicolumn{1}{c}{$\rm Rate_{\rm SI/SII}$} & S \\
No. & \multicolumn{1}{c}{($\mbox{ct} * \mbox{ks}^{-1}$)} & \multicolumn{1}{c}{($\mbox{ct} * \mbox{ks}^{-1}$)} & \\ \hline}
\tabletail{\hline} 
\scriptsize
\begin{supertabular}{crrrrr}
	RXJ0038.4+4012 & $  12.61 \pm   0.66 $ & $   9.03  \pm   0.78 $ & $   1.40 \pm   0.14 $ &   3.51 \\
	RXJ0040.7+3959 & $            < 2.81 $ & $   5.54  \pm   1.01 $ & $            < 0.51 $ &      T \\
	RXJ0041.1+4002 & $   1.96 \pm   0.64 $ & $   5.57  \pm   1.01 $ & $   0.35 \pm   0.13 $ &   3.04 \\
	RXJ0041.5+4105 & $            < 1.27 $ & $  12.36  \pm   0.69 $ & $            < 0.10 $ &      T \\
	RXJ0041.6+4101 & $   1.49 \pm   0.31 $ & $   3.10  \pm   0.40 $ & $   0.48 \pm   0.12 $ &   3.20 \\
	RXJ0041.8+4015 & $   3.18 \pm   0.58 $ & $   7.02  \pm   1.04 $ & $   0.45 \pm   0.11 $ &   3.24 \\
	RXJ0041.8+4021 & $  60.52 \pm   1.22 $ & $  90.56  \pm   2.85 $ & $   0.67 \pm   0.02 $ &   9.68 \\
	RXJ0041.8+4101 & $  11.73 \pm   0.71 $ & $   6.25  \pm   0.51 $ & $   1.88 \pm   0.19 $ &   6.28 \\
	RXJ0041.8+4122 & $   6.11 \pm   0.58 $ & $   2.69  \pm   0.43 $ & $   2.27 \pm   0.42 $ &   4.75 \\
	RXJ0042.1+4110 & $   4.20 \pm   0.46 $ & $   6.83  \pm   0.65 $ & $   0.62 \pm   0.09 $ &   3.28 \\
	RXJ0042.1+4118 & $  13.85 \pm   0.79 $ & $  35.01  \pm   1.21 $ & $   0.40 \pm   0.03 $ &  14.68 \\
	RXJ0042.2+4039 & $   3.90 \pm   0.39 $ & $   8.12  \pm   0.71 $ & $   0.48 \pm   0.06 $ &   5.24 \\
	RXJ0042.2+4055 & $  10.13 \pm   0.69 $ & $   6.59  \pm   0.52 $ & $   1.54 \pm   0.16 $ &   4.10 \\
	RXJ0042.2+4101 & $  35.60 \pm   1.18 $ & $  29.59  \pm   1.02 $ & $   1.20 \pm   0.06 $ &   3.85 \\
	RXJ0042.2+4112 & $  24.95 \pm   1.02 $ & $  33.74  \pm   0.95 $ & $   0.74 \pm   0.04 $ &   6.33 \\
	RXJ0042.2+4118 & $  11.31 \pm   0.73 $ & $  36.25  \pm   1.21 $ & $   0.31 \pm   0.02 $ &  17.64 \\
	RXJ0042.3+4113 & $  18.74 \pm   0.87 $ & $  66.18  \pm   0.96 $ & $   0.28 \pm   0.01 $ &  36.58 \\
	RXJ0042.4+4104 & $   6.95 \pm   0.56 $ & $  17.85  \pm   0.83 $ & $   0.39 \pm   0.04 $ &  10.87 \\
	RXJ0042.4+4112 & $  22.09 \pm   0.94 $ & $  44.15  \pm   0.76 $ & $   0.50 \pm   0.02 $ &  18.34 \\
	RXJ0042.5+4048 & $   1.69 \pm   0.32 $ & $   3.91  \pm   0.46 $ & $   0.43 \pm   0.10 $ &   3.97 \\
	RXJ0042.6+4052 & $ 121.17 \pm   1.99 $ & $  58.12  \pm   1.50 $ & $   2.08 \pm   0.06 $ &  25.28 \\
	RXJ0042.8+4125 & $  14.75 \pm   0.84 $ & $  21.27  \pm   0.98 $ & $   0.69 \pm   0.05 $ &   5.05 \\
	RXJ0042.9+4146 & $   3.55 \pm   0.53 $ & $   7.64  \pm   0.72 $ & $   0.46 \pm   0.08 $ &   4.57 \\
	RXJ0043.1+4048 & $            < 2.27 $ & $   5.53  \pm   0.62 $ & $            < 0.41 $ &      T \\
	RXJ0043.1+4112 & $   3.06 \pm   0.41 $ & $   5.63  \pm   0.56 $ & $   0.54 \pm   0.09 $ &   3.69 \\
	RXJ0043.1+4118 & $   5.58 \pm   0.57 $ & $  28.17  \pm   0.91 $ & $   0.20 \pm   0.02 $ &  21.05 \\
	RXJ0043.3+4120 & $   6.74 \pm   0.62 $ & $  10.07  \pm   0.91 $ & $   0.67 \pm   0.09 $ &   3.02 \\
	RXJ0043.4+4118 & $   6.86 \pm   0.62 $ & $  11.96  \pm   0.76 $ & $   0.57 \pm   0.06 $ &   5.20 \\
	RXJ0043.4+4126 & $            < 1.50 $ & $   5.87  \pm   0.52 $ & $            < 0.26 $ &      T \\
	RXJ0043.7+4124 & $            < 1.53 $ & $   3.72  \pm   0.50 $ & $            < 0.41 $ &      T \\
	RXJ0043.7+4136 & $   7.04 \pm   0.56 $ & $   2.16  \pm   0.36 $ & $   3.26 \pm   0.60 $ &   7.35 \\
	RXJ0043.9+4122 & $   4.82 \pm   0.48 $ & $   2.07  \pm   0.36 $ & $   2.33 \pm   0.47 $ &   4.56 \\
	RXJ0044.3+4145 & $   1.17 \pm   0.35 $ & $   3.07  \pm   0.42 $ & $   0.38 \pm   0.12 $ &   3.51 \\
	RXJ0044.4+4121 & $  29.77 \pm   1.12 $ & $  25.10  \pm   1.02 $ & $   1.19 \pm   0.07 $ &   3.08 \\
	RXJ0044.8+4225 & $            < 2.74 $ & $   4.85  \pm   1.11 $ & $            < 0.56 $ &      T \\
	RXJ0045.6+4208 & $  19.18 \pm   0.92 $ & $  25.83  \pm   1.11 $ & $   0.74 \pm   0.05 $ &   4.63 \\
	RXJ0045.7+4139 & $ 147.96 \pm   1.91 $ & $ 134.43  \pm   1.96 $ & $   1.10 \pm   0.02 $ &   4.95 \\
	RXJ0046.4+4201 & $  29.58 \pm   1.14 $ & $  37.64  \pm   1.15 $ & $   0.79 \pm   0.04 $ &   4.99 \\
	RXJ0046.4+4204 & $  20.61 \pm   0.99 $ & $  33.17  \pm   1.16 $ & $   0.62 \pm   0.04 $ &   8.24 \\
	RXJ0047.4+4152 & $            < 2.08 $ & $   3.35  \pm   0.54 $ & $            < 0.62 $ &      T \\                  
	RXJ0047.8+4142 & $            < 2.70 $ & $   7.02  \pm   1.15 $ & $            < 0.38 $ &      T \\
	RXJ0048.4+4157 & $  46.28 \pm   1.64 $ & $  35.30  \pm   2.16 $ & $   1.31 \pm   0.09 $ &   4.05 \\
\end{supertabular}
\normalsize
\end{table}

With these criteria, 34 possible long term variable sources were found, as
listed in Table \ref{varSurveysList}. Column (1) gives the ROSAT RXJ--number
of the source, column (2) and (3) list the count rate in the B--band
determined from the data of the first and second surveys respectively, column
(4) gives the ratio in count rate between the first and second survey, and
column (5) gives the value of the significance parameter, following eq.\
(\ref{S}).

Additionally, Table \ref{varSurveysList} contains possible transients, marked
with a `T' in column (5). For this, the sensitive flux limit was determined
within the survey in which the source was {\em not} found, using the source
position from the other survey (i.e. where the source {\em was} detected).
If this value was below the count rate minus the $1\sigma$ count rate error
determined from the survey where the source was found, then this source was
considered as a possible transient. To prevent false diagnoses being made, the
same criteria as above for the variable source search were applied except for
criterion (2) which was dropped, and criterion (3), which was substituted as
just described. With this, no transients were found which could be seen only
in the first survey but not in the second. This is mainly due to the exclusion
of regions near the PSPC support structure within the first survey which
results in a reduction in area and may have removed a few transient candidates
from our (conservative) list. Additionally, the second survey with its
homogeneous exposure is more sensitive in the outer region of \object{M31}
than the first survey. As a consequence of these two effects, we found 8
transients which were seen in the second survey but not in the first. Because
of the very different conditions of both surveys (mainly the influence of the
PSPC support structure in the first survey), we desist from a quantitative
analysis of a transient rate and its comparison with expected theoretical
values.

If we readopt criterion (2) in a slightly changed form, that the upper flux
limit for transient sources also represents the error in flux, we would come
up with values for the significance parameter always below our threshold of 3
except for source RXJ0041.5+4105 where $S = 7.67$. Here, we could quote source
RXJ0041.5+4105 as a strong candidate for a transient, whereas all the others
must be considered as weak candidates.

Some words concerning ROSAT source RXJ0040.2+4034: In Sect.\
\ref{VarEinstein}, from a comparison with the {\it Einstein} source list of
TF, we have indicated this source as a possible transient. If the increase in
flux between the {\it Einstein} observations and the first ROSAT survey is
based on a short--time outburst of this transient source, we would expect this
source to appear much fainter during the second PSPC survey or even disappear.
Actually, with the criteria applied to merge both source lists as described in
Sect.\ \ref{sourcedescript}, the source seemed to disappear, as no
correlating source could be found within the second survey. Nevertheless, a
visual inspection suggested an identification of ROSAT source RXJ0040.2+4034, 
only found within the first survey (source \#69, Table 5 in S97), with ROSAT
source RXJ0040.2+4033,  only detected within the second survey. Both sources
are listed separately in Table \ref{sourcelist}. Under the assumption that
these two sources are the same source we note a large decrease in count rate
(by over a factor of 40) between the first and second surveys.
This would tie in with the possible transient nature
of this source. On the other hand, the fact that these two sources are 
separated by $57 \arcsec$ and are both good quality detections argues
against this treatement. We therefore list both sources as individual
sources in our list.

\section{Total luminosity and diffuse emission}
\label{Luminosity}

From the first ROSAT PSPC survey of \object{M31} we had already derived
quantities for the total luminosity of \object{M31} and a possible gaseous
component (S97). Because the first survey had to be corrected for several
caveats such as the dominant influence of the PSPC support structure, the
inhomogeneous exposure, and the rapid decrease of sensitivity from the centre
of \object{M31} to the outer regions, we improved the determination of the
total luminosity and diffuse component with the data from the much more
homogeneous second survey. One of the big advantages of the second survey is
its more or less constant exposure and therefore constant flux limit over the
whole $D_{\rm 25}$-area of the galaxy. This allows an improved determination of
the background around \object{M31} and, as a consequence, a more reliable flux
determination of components within \object{M31}. Furthermore, it reduces
systematical errors in the case of large scale analysis, as discussed in this
section. The following description has some overlap with procedures already
described in S97, but we decided to briefly summarise them here for
completeness.

In this section we will use the term ``diffuse com\-po\-nent'' to mean the sum
of the emission from a truly diffuse (gaseous) emitter and from unresolved
point sources. We will refer to ``total emission'' as the sum of the diffuse
component and the emission from resolved point sources.

As already described in Sect.\ \ref{Data preparation}, all the data have been
cleaned of contamination by solar scattered X-rays and particle background.
The resulting photon event files remain contaminated by these components, but
only to less than 1\% in each pointing.
This is up to ten times better than in the worst case of the first PSPC
survey. For the analysis in this section, the data were binned into an image
with a $30\arcsec \times 30\arcsec$ pixel size. For the determination of count
rates within the $D_{\rm 25}$-area of \object{M31}, the merged inner regions of
the PSPC with $20\arcmin$ radius have been used, whereas for the outer area
around \object{M31}, a merge of the total photon event files has been used. The
resulting images were divided by exposure maps with the same pixel size to
obtain count rate images corrected for the effects of the rib structure,
vignetting and dead time. These exposure maps were calculated in the following
manner: the B-band was divided into 10 energy slices for which EXSAS provides
instrument maps for the PSPC detector response. Together with the photon event
files, exposure maps for each of these energy slices were created, considering
also dead time effects.  A weighted addition of these single exposure maps
yields the final exposure maps. The pulse height spectra in the 10 energy
slices of the photon event files were used as the weighting factors.

From the image of the merged inner PSPC regions we derived count rates for
the bulge (1 kpc around the centre) and the \object{M31} disk region (i.e.
outside the bulge up to the $D_{\rm 25}$-ellipse). ``Background count rates'' were
taken from the image of the merged total PSPC FOV and within an area far
outside and around the $D_{\rm 25}$ ellipse of \object{M31} -- explicitly the area
between the ellipse with major and minor axes $0.15\degr$ larger than the
$D_{\rm 25}$ ellipse of \object{M31} and the ellipse $0.30\degr$ larger. Sources
within this area were cut out to a radius of three times the PSF at the source
position. With this, we derived count rates for the bulge, disk, and
``background'' of $(46.86\pm2.5)$, $(4.278\pm0.04)$, and $(3.311\pm0.038)
\mbox{ ct s}^{-1}\mbox{ deg}^{-2}$ respectively, in the broad (0.1 - 2.0 keV)
energy band.

Considering the bulge, a subtraction of the background count rate and a
multiplication with the bulge area of $0.026 \mbox{ deg}^2$ yields
$(1.132\pm0.065) \mbox{ ct s}^{-1}$. Applying a power law with $\Gamma = -2.0$
for the spectral model and a galactic foreground absorption of $N_{\rm H} =
6 \times 10^{20} \mbox{ cm}^{-2}$ yields $(2.88\pm0.17) \times 10^{-11} \mbox{
erg cm}^{-2} \mbox{ s}^{-1}$ for the total flux of the bulge region, which
corresponds to a luminosity of $\sim1.6 \times 10^{39} \mbox{ erg s}^{-1}$,
assuming a distance of 690 kpc to \object{M31}. A summation over the count
rates of all 22 bulge sources detected in the second PSPC survey data in this
area initially yields $(2.78\pm0.02) \mbox{ ct s}^{-1}$. This is much higher
than the total emission derived above. The reason is the way the source
detection algorithm works. In highly confused regions it tends to overestimate
the count rate of each source due to overlapping of the photon extraction
circles of neighbouring sources. By determining the individual extraction
radii the detection algorithm has used, and the amount of overlapping area
under the assumption of a gaussian PSF for the instrumentation, we can
globally correct for this effect. With this, we obtain $(0.893\pm0.006) \mbox{
ct s}^{-1}$ for the resolved emission of the bulge. A comparison with the
above derived total emission uncovers an unresolved component of
$(0.239\pm0.065) \mbox{ ct s}^{-1}$. Assuming that this component completely
originates from thermal emission of hot gas, and applying a
spectral model for an optically-thin thermal plasma (MEKAL)
with $kT = 0.35 \mbox{ keV}$ (as determined from XMM-Newton
observations, e.g. see Shirey et al.\ \cite{Shi01}) and a galactic
foreground absorption of $N_{\rm H} = 6 \times 10^{20} \mbox{ cm}^{-2}$, we
derive $(3.4\pm0.9) \times 10^{-12} \mbox{ erg cm}^{-2} \mbox{ s}^{-1}$ for a
diffuse X-ray flux. For a distance of 690 kpc to \object{M31}, this corresponds
to a luminosity of $(2.0\pm0.5) \times 10^{38} \mbox{ erg s}^{-1}$ and would
indicate a gas mass of $(1.0\pm0.3) \times 10^6 \,\mbox{M}_{\sun}$, assuming
the gas fills uniformly the bulge region, a sphere with 1 kpc radius (using
the power per unit emission integral as a function of temperature for a low
density plasma reported by Kato \cite{Kat76}). Because a luminosity function
derived from the detected sources in the heavily confused bulge region would
be very uncertain, we cannot trust any estimation of the emission from
non-detected sources below our detection threshold by extrapolating such a
luminosity function. As a consequence, the above derived luminosity (and gas
mass) of the diffuse emission must be considered as an upper limit.

Considering the disk, a subtraction of the background count rate and a
multiplication with the disk area of $2.6 \mbox{ deg}^2$ yields $(1.68\pm0.14)
\mbox{ ct s}^{-1}$. A summation of the count rates of all the sources detected
in the disk within the second PSPC survey data yields $(2.06\pm0.31) \mbox{ ct
s}^{-1}$. Here no correction had to be applied, as no important source
confusion exists. This value is slightly higher than the one derived from the
total emission. It may indicate a possible diffuse absorption of background
photons by \object{M31}. Although both derived count rates are comparable
within their $1\sigma$ errors, this is an effect of the integral consideration
of the whole disk. A division into several annular regions indicates an
absorption at the $1\sigma$ significance level in some of these regions. A
more detailed report will be the subject of a future paper. In the following
discussion, we neglect a possible (slight) absorption in the \object{M31}
disk.

As already mentioned in Sect. \ref{corall}, a fair number of the detected
sources do not belong to \object{M31}, but are foreground sources or
background sources shining through the galaxy. Therefore, the derived flux of
all the resolved disk sources mentioned above (or the sum of the flux in the
disk area) cannot be used for a determination of the total X-ray luminosity of
the disk of \object{M31}. Following the procedure described in S97 we use the
there derived logN-logS distribution for sources truly belonging to
\object{M31} (from a statistical point of view). We come up with
$(1.26\pm0.20) \mbox{ ct s}^{-1}$ for the resulting count rate, or a total flux
of $(1.7\pm0.3) \times 10^{-11} \mbox{ erg cm}^{-2} \mbox{ s}^{-1}$ for the
disk of \object{M31} (using the above spectral model). This
corresponds to a total luminosity of $(1.8\pm0.3) \times 10^{39} \mbox{ erg
s}^{-1}$.

All together, applying a power law spectral model with photon index $\Gamma =
-2.0$ and a galactic foreground absorption of $N_{\rm H} = 6 \times 10^{20}
\mbox{ cm}^{-2}$, we obtain for the total (0.1$-$2.0 keV) luminosity of
\object{M31}, $(3.4\pm0.3) \times 10^{39} \mbox{ erg s}^{-1}$, approximately
equally distributed between the bulge and disk.

\vspace{2ex}
{\it Comparison with earlier results:}

A comparison with the results derived from the first ROSAT PSPC survey of
\object{M31} (S97) uncovered a difference in the bulge luminosities. For the
total emission as well as for the sum of the resolved flux of detected sources
we determined slightly higher values from the second PSPC survey data.
Although the difference in significance for the total emission is less than
$1.5\sigma$, we decided to take the new value from the second survey as the
better one due to the above mentioned reasons. Because the flux of the
resolved emission increased approximately by the same (small) amount we would
obtain nearly the same value for a possible gaseous component in the bulge
of \object{M31} as previously derived from the first survey data when applying
the same spectral model as used in S97 (now
$(4.4\pm1.2) \times 10^{-12} \mbox{ erg cm}^{-2} \mbox{ s}^{-1}$, compared to
$(4.6\pm1.1) \times 10^{-12} \mbox{ erg cm}^{-2} \mbox{ s}^{-1}$ in S97).
It shows, that the change of the here newly given value
($(3.4\pm0.9) \times 10^{-12} \mbox{ erg cm}^{-2} \mbox{ s}^{-1}$)
is mainly due to the new spectral model used (optically-thin thermal
plasma with $kT = 0.35 \mbox{ keV}$), which we adopted from recent
results of XMM-Newton observations (Shirey et al. \cite{Shi01}).
With this, the ROSAT derived diffuse luminosity within $5\arcmin$ of the nucleus of M31 
is comparable to the luminosity found for the same bulge area with the 
Chandra (Garcia et al. \cite{Gar00}) and the XMM-Newton observation (Shirey et al. 
\cite{Shi01}). It is commonly assumed that the hot component of the interstellar medium
(ISM) is created by 
winds from massive young stars and supernova explosions in star-forming regions.
The diffuse emission from the hot ISM 
in M31 is less pronounced than that detected from the
inner spiral arms in the neighboring Local Group galaxy M33. For this galaxy
ROSAT HRI (Shulman \& Bregman \cite{Shu94}) and PSPC observations (Long et al. \cite{Lon96})
show diffuse emission with a luminosity of about $10^{39} \mbox{ erg s}^{-1}$ that traces the
spiral arms within $15\arcmin$ of the nucleus and has a temperature of $kT = 0.4 \mbox{ keV}$.
Galaxies with high star-forming activity may be even brighter in diffuse
X-rays by factors of more than 10 (see e.g. Read et al. \cite{Rea97}, Vogler \& 
Pietsch \cite{Vog99a}). The low diffuse X-ray luminosity in M31 therefore supports 
the view that the galaxy is in a phase of low star-forming activity.

For the determination of the disk luminosity we adopted the procedure from our
previous calculations used in the first survey. Hence, we obtained the
same results. Also the considerations concerning the normalized luminosity distribution
of the discrete X-ray sources in the disk of M31 are still valid (see S97).
A comparison with the luminosity distributions (normalized to bulge luminosity)
of other nearby spiral galaxies like M33, M51, M83, M100, M101, NGC253, NGC1566, NGC4258,
NGC4559, NGC4565, and NGC4631 (see Vogler \& Pietsch \cite{Vog99b})
shows no significant differencies in shape and
reveals the distribution of M31 as being typical for this class of galaxy.
However, we do not find super-luminous sources (SLS) above several times 
$10^{38} \mbox{erg s}^{-1}$, as is also the case in M33 and NGC253, but
not for the other (star-forming) galaxies mentioned above. Although NGC253 is a (bulge)
star-forming galaxy it shows no SLSs in its disk population.
Therefore it is difficult to interpret the absence of SLSs in M31, but 
it perhaps tends to show that M31 is not in a star-forming phase.

The discussion of the comparison of our results with those obtained from the
{\it Einstein} observatory and reported by TF also changes slightly under the
transition from the first to the second PSPC survey. For the total
luminosity of \object{M31}, TF found a value of $\sim3 \times 10^{39} \mbox{
erg s}^{-1}$. To compare with our values, one has to take into account the
different spectral models, energy ranges, and especially the different fields
of \object{M31} investigated.  TF derived the luminosities from the {\it
Einstein} data by applying a thermal bremsstrahlung spectrum in the energy
band 0.2 keV - 4.0 keV with $kT = 5 \mbox{ keV}$ and $N_{\rm H} = 7 \times
10^{20} \mbox{ cm}^{-2}$.  They integrated the count rates within an ellipse
of $\sim2.5\degr \times 1.0\degr$ which is a bit smaller than the $D_{\rm 25}$
ellipse we used for our calculations. A conversion of our results to the
spectral model and reduced area of TF yields for the total luminosity
$(3.3\pm0.3) \times 10^{39} \mbox{ erg s}^{-1}$. The $1\sigma$ agreement with
the value reported by TF, however, is somewhat coincidental: while our
observations covered the whole galaxy, those of TF did not. On the other hand,
TF did not correct for background sources.

Comparing the total luminosity of the bulge region, TF reported $\sim1.5
\times 10^{39} \mbox{ erg s}^{-1}$, which is in agreement with our value of
$1.6 \times 10^{39} \mbox{ erg s}^{-1}$ (in this case the effect of the
different assumed spectral models is below the errors and therefore
negligible). In contrast, for the disk alone we found a somewhat higher
luminosity ($(1.8\pm0.3) \times 10^{39} \mbox{ erg s}^{-1}$) than TF ($\sim1.5
\times 10^{39} \mbox{ erg s}^{-1}$), though there is still a $1\sigma$
agreement. Considering the fact that TF did not describe the errors and
furthermore did not explicitly quote the values for the bulge and disk
emission, but simply mentioned that ``the emission is roughly equally 
divided between the bulge and the disk'', as well as their neglecting to compensate for
background/foreground sources, we desist from a more quantitative comparison,
noting that the agreement is surprisingly good. Our results tend to show that
TF determined the disk luminosity too low and with it, the total luminosity of
\object{M31}. With the improved capabilities of ROSAT, the complete coverage
of the total galaxy, and our considerations of statistical errors, we were able
to clarify the luminosities in \object{M31} at a more reliable level.

As already mentioned, the second survey data did not (significantly) change
the results concerning a possible diffuse emission component in the bulge
region (from $(2.6\pm0.6) \times 10^{38} \mbox{ erg s}^{-1}$ to $(2.5\pm0.7)
\times 10^{38} \mbox{ erg s}^{-1}$, when using the spectral model of S97). 
The exhaustive discussion of the
comparison with the value reported by TF ($\sim3.8 \times 10^{38} \mbox{ erg
s}^{-1}$) and the reasons for the difference have already been undertaken in
S97, and are still valid.

\section{Summary and conclusions}
\label{Summary}

The second pointed ROSAT PSPC survey of \object{M31} has extended
our knowledge concerning the X-ray nature of this spiral galaxy
beyond that already derived from the first survey described
in S97 (Supper et al.\ \cite{Sup97}). Merging the two point source
lists of the two surveys led to a total of 560 X-ray sources 
in the $\sim10.7 \mbox{ deg}^2$ \object{M31} FOV, 
31 located in the very confused bulge region.
Their luminosities range from $4 \times 10^{35} \mbox{ erg s}^{-1}$
to $4 \times 10^{38} \mbox{ erg s}^{-1}$, assuming
a distance of 690 kpc to \object{M31}. Of these sources, 55 have
been identified with known foreground stars, 33 with
globular clusters, 16 with supernova remnants, and 
10 correlate with known background objects such as background
galaxies. None of our \object{M31} sources could be assigned to 
known novae. A comparison with the {\it Einstein}
source list reported by TF confirms 69 {\it Einstein} sources.
The much improved homogeneity of the second PSPC survey compared
with the first and the resulting fewer problems with
the PSPC support structure, allowed better flux determinations
for a couple of sources. Combined with the higher positional 
precision in some regions, the list of variable sources
when compared with the reported {\it Einstein} source fluxes
could be restricted to 11 candidates, and 7 transients were discovered. 
Comparisons of the {\it Einstein} source list with the two ROSAT survey 
source lists separately, may yield up to 10 transients. 
Finally, of the 60 sources reported by TF outside the
heavily confused bulge region, we could confirm 54, 
or 90\% of these sources. In total, 
39 {\it Einstein} sources could not be confirmed, while
491 new sources were found with ROSAT. 

Comparing the first and second 
PSPC surveys of \object{M31}, 34 possible long term variable sources 
and 8 possible transients (with some overlap with the
transients obtained from the comparison with the
{\it Einstein} detected sources) are reported. 

For the bulge region, we can give an upper limit to the diffuse component
luminosity of $(2.0\pm0.5) \times 10^{38} \mbox{ erg s}^{-1}$ when using
an optically-thin thermal plasma (MEKAL) with $kT = 0.35 \mbox{ keV}$
for the spectral model. This is a factor of
$\sim1.5$ lower than the value reported by TF (after transforming
to their spectral model). If we assume this luminosity as
completely originating from hot gas within the bulge
region, this would indicate a gas mass upper limit of $(1.0\pm0.3) \times 10^6
\,\mbox{M}_{\sun}$.  For the total (0.1 -- $2.0 \mbox{ keV}$) luminosity of
\object{M31}, we obtain $(3.4\pm0.3) \times 10^{39} \mbox{ erg s}^{-1}$, for
the bulge alone $1.6 \times 10^{39} \mbox{ erg s}^{-1}$ and for the disk
$(1.8\pm0.3) \times 10^{39} \mbox{ erg s}^{-1}$. With these improved values,
we find an equal distribution of luminosity between the bulge and disk, in
agreement with TF, but a higher value for the total luminosity than reported in
TF.

Several results from the first PSPC survey of M31 reported in S97 have
not been significantly altered by the inclusion of the second survey
data and remain valid.  These include: 1) the integral luminosity
distribution of the globular cluster sources and its comparison to that
of the Milky Way, 2) the statistical estimation of the fraction of
background and foreground sources among the detected X-ray sources,
and 3) the spectral analysis of the brightest sources.

\begin{acknowledgements}
We thank the MPE ROSAT group for their support. Parts of our analysis
used the SASS and EXSAS data analysis software.
The authors would like to thank the referee for carefully 
reading this paper and making useful comments and suggestions.
We thank NASA and the Centre de Donn\'ees
astronomiques for the online access to their data bases.
WHGL is grateful for support from NASA.\newline
RS remembers J.\ v.\ Paradijs as an always gentle and
helpful excellent scientist who sadly passed
away well ahead of his time -- a great loss for the 
astronomical community.\newline
The ROSAT
project has been supported by the Bundesministerium f\"ur Bildung
und Forschung (BMBF/DLR) and the Max-Planck-Gesellschaft (MPG).
\end{acknowledgements}


\onecolumn

\begin{table*}[h]
\caption{\label{journal}
         Log of the 94 single observations forming the second ROSAT PSPC 
         survey of \object{M31}. Pointing numbers ending with ``-1'' or ``-2'' 
         indicate follow-up observations.}
\end{table*}
\tablehead{\hline
Pointing  & \multicolumn{2}{c}{Date} & \multicolumn{3}{l}{R.A. (J2000)$^1$} &\ & \multicolumn{3}{l}{
				Dec. (J2000)$^1$} & Exposure \\ \cline{4-6}\cline{8-10}
			 &       &            & (h) & (m) & (s) & & ($\degr$) & ($'$)
			 & ($''$) & (s)\\ \hline \hline}
\tabletail{\hline}
\begin{flushleft}
\begin{supertabular}{ll\hs rrrrrrrrr}
WG600296P     & 25.-26. July & 1992  &  0 & 37 & 43.2 & & 40 & 23 & 24 &  2\,512 \\
WG600297P     & 25.-25. July & 1992  &  0 & 38 &  9.6 & & 40 & 29 & 24 &  2\,536 \\
WG600316P     & 21.-21. July & 1992  &  0 & 38 & 26.3 & & 40 & 17 & 24 &  2\,672 \\
WG600298P     & 05.-05. Aug. & 1992  &  0 & 38 & 33.5 & & 40 & 36 & 00 &  2\,640 \\
WG600317P     & 20.-20. July & 1992  &  0 & 38 & 52.7 & & 40 & 23 & 24 &  2\,616 \\
WG600299P     & 05.-05. Aug. & 1992  &  0 & 39 &  0.0 & & 40 & 42 &  0 &  1\,576 \\
WG600299P-1   & 01.-11. Jan. & 1993  &  0 & 39 &  0.0 & & 40 & 42 &  0 &  1\,448 \\
WG600336P     & 29.-30. July & 1992  &  0 & 39 & 12.0 & & 40 & 11 & 24 &  2\,752 \\
WG600318P     & 06.-06. Aug. & 1992  &  0 & 39 & 16.7 & & 40 & 30 &  0 &  1\,416 \\
WG600318P-1   & 31.-31. Dec. & 1992  &  0 & 39 & 16.7 & & 40 & 30 &  0 &  1\,312 \\
WG600300P     & 05.-05. Aug. & 1992  &  0 & 39 & 24.0 & & 40 & 48 & 36 &  2\,456 \\
WG600337P     & 30.-30. Dec. & 1992  &  0 & 39 & 36.0 & & 40 & 17 & 24 &  2\,096 \\
WG600319P     & 06.-06. Aug. & 1992  &  0 & 39 & 43.2 & & 40 & 36 & 00 &  1\,816 \\
WG600301P     & 22.-22. July & 1992  &  0 & 39 & 48.0 & & 40 & 54 & 36 &  2\,688 \\
WG600356P     & 03.-04. Aug. & 1992  &  0 & 39 & 55.2 & & 40 &  5 & 24 &  1\,976 \\
WG600338P     & 03.-03. Aug. & 1992  &  0 & 40 &  2.4 & & 40 & 24 &  0 &  2\,168 \\
WG600320P-1   & 01.-01. July & 1993  &  0 & 40 &  7.2 & & 40 & 42 & 36 &  2\,744 \\
WG600302P     & 30.-31. Dec. & 1992  &  0 & 40 & 14.3 & & 41 &  1 & 12 &  2\,568 \\
WG600357P     & 30.-30. Dec. & 1992  &  0 & 40 & 21.6 & & 40 & 11 & 24 &  2\,720 \\
WG600339P     & 25.-25. July & 1992  &  0 & 40 & 26.3 & & 40 & 30 &  0 &  2\,840 \\
WG600321P     & 29.-30. July & 1992  &  0 & 40 & 33.5 & & 40 & 48 & 36 &  2\,848 \\
WG600303P     & 27.-28. July & 1992  &  0 & 40 & 38.4 & & 41 &  7 & 12 &  2\,416 \\
WG600358P     & 26.-26. July & 1992  &  0 & 40 & 45.5 & & 40 & 18 & 00 &  2\,864 \\
WG600340P     & 06.-06. Aug. & 1992  &  0 & 40 & 52.7 & & 40 & 36 & 36 &  2\,560 \\
WG600322P     & 06.-06. Aug. & 1992  &  0 & 40 & 57.5 & & 40 & 55 & 12 &  2\,584 \\
WG600304P     & 31.-31. Dec. & 1992  &  0 & 41 &  4.8 & & 41 & 13 & 48 &  1\,872 \\
WG600359P     & 06.-07. Aug. & 1992  &  0 & 41 & 12.0 & & 40 & 24 &  0 &  2\,600 \\
WG600341P     & 23.-23. July & 1992  &  0 & 41 & 16.7 & & 40 & 42 & 36 &  2\,712 \\
WG600323P     & 05.-06. Aug. & 1992  &  0 & 41 & 24.0 & & 41 &  1 & 12 &  2\,560 \\
WG600305P     & 06.-07. Aug. & 1992  &  0 & 41 & 28.7 & & 41 & 19 & 48 &  2\,216 \\
WG600360P     & 23.-24. July & 1992  &  0 & 41 & 36.0 & & 40 & 30 & 36 &  2\,864 \\
WG600342P     & 05.-05. Aug. & 1992  &  0 & 41 & 43.2 & & 40 & 49 & 12 &  3\,416 \\
WG600324P     & 07.-07. Jan. & 1993  &  0 & 41 & 48.0 & & 41 &  7 & 48 &  2\,960 \\
WG600306P     & 07.-08. Aug. & 1992  &  0 & 41 & 55.2 & & 41 & 26 & 24 &  2\,424 \\
WG600361P     & 05.-06. Aug. & 1992  &  0 & 42 &  0.0 & & 40 & 36 & 36 &  2\,744 \\
WG600343P     & 04.-04. Jan. & 1993  &  0 & 42 &  7.1 & & 40 & 55 & 12 &  2\,744 \\
WG600325P     & 02.-02. Jan. & 1993  &  0 & 42 & 14.3 & & 41 & 13 & 48 &  2\,848 \\
WG600307P     & 01.-01. Jan. & 1993  &  0 & 42 & 19.2 & & 41 & 32 & 24 &   720 \\
WG600307P-1   & 05.-09. July & 1993  &  0 & 42 & 19.2 & & 41 & 32 & 24 &  2\,032 \\
WG600362P     & 06.-06. Aug. & 1992  &  0 & 42 & 26.3 & & 40 & 43 & 12 &  2\,448 \\
WG600344P     & 07.-07. Aug. & 1992  &  0 & 42 & 31.2 & & 41 &  1 & 48 &  1\,744 \\
WG600326P     & 07.-07. Aug. & 1992  &  0 & 42 & 38.4 & & 41 & 20 & 24 &  1\,704 \\
WG600326P-2   & 18.-18. July & 1993  &  0 & 42 & 38.4 & & 41 & 20 & 24 &   696 \\
WG600308P     & 07.-07. Aug. & 1992  &  0 & 42 & 45.5 & & 41 & 39 &  0 &  1\,512 \\
WG600308P-1   & 03.-11. Jan. & 1993  &  0 & 42 & 45.5 & & 41 & 39 &  0 &  1\,768 \\
WG600363P     & 03.-03. Jan. & 1993  &  0 & 42 & 50.4 & & 40 & 49 & 12 &  1\,944 \\
WG600345P     & 01.-01. Jan. & 1993  &  0 & 42 & 57.5 & & 41 &  7 & 48 &  1\,592 \\
WG600345P-1   & 22.-22. July & 1993  &  0 & 42 & 57.6 & & 41 &  7 & 48 &  1\,712 \\
WG600327P     & 07.-07. Aug. & 1992  &  0 & 43 &  4.8 & & 41 & 26 & 24 &  2\,112 \\
WG600309P     & 07.-08. Aug. & 1992  &  0 & 43 &  9.6 & & 41 & 45 &  0 &  2\,576 \\
WG600364P     & 07.-07. Aug. & 1992  &  0 & 43 & 16.7 & & 40 & 55 & 48 &   416 \\
WG600364P-2   & 04.-04. July & 1993  &  0 & 43 & 16.8 & & 40 & 55 & 48 &  3\,296 \\
WG600346P     & 06.-06. Aug. & 1992  &  0 & 43 & 21.6 & & 41 & 14 & 24 &  1\,968 \\
WG600328P     & 02.-02. Jan. & 1993  &  0 & 43 & 28.7 & & 41 & 33 & 00 &  2\,320 \\
WG600310P     & 05.-05. Jan. & 1993  &  0 & 43 & 36.0 & & 41 & 51 & 36 &  2\,720 \\
WG600365P     & 02.-02. Jan. & 1993  &  0 & 43 & 40.7 & & 41 &  1 & 48 &  3\,144 \\
WG600347P     & 03.-03. Jan. & 1993  &  0 & 43 & 48.0 & & 41 & 20 & 24 &  2\,936 \\
WG600329P     & 08.-08. Aug. & 1992  &  0 & 43 & 52.7 & & 41 & 39 &  0 &  1\,760 \\
WG600311P     & 08.-08. Aug. & 1992  &  0 & 44 &  0.0 & & 41 & 58 & 12 &  1\,352 \\
WG600311P-1   & 01.-01. Jan. & 1993  &  0 & 44 &  0.0 & & 41 & 58 & 12 &  1\,240 \\
WG600366P     & 07.-07. Aug. & 1992  &  0 & 44 &  7.1 & & 41 &  8 & 24 &  1\,776 \\
WG600348P     & 31.-31. Dec. & 1992  &  0 & 44 & 12.0 & & 41 & 27 &  0 &  1\,696 \\
WG600348P-1   & 22.-22. July & 1993  &  0 & 44 & 12.0 & & 41 & 27 &  0 &  1\,800 \\
WG600330P     & 08.-08. Aug. & 1992  &  0 & 44 & 19.2 & & 41 & 45 & 36 &  1\,624 \\
WG600330P-1   & 14.-14. Jan. & 1993  &  0 & 44 & 19.2 & & 41 & 45 & 36 &   944 \\
WG600312P     & 08.-08. Aug. & 1992  &  0 & 44 & 24.0 & & 42 &  4 & 12 &  2\,416 \\
WG600367P     & 04.-04. Jan. & 1993  &  0 & 44 & 31.2 & & 41 & 14 & 24 &  2\,624 \\
WG600349P     & 06.-06. Jan. & 1993  &  0 & 44 & 38.4 & & 41 & 33 & 00 &  2\,824 \\
WG600331P     & 04.-04. Jan. & 1993  &  0 & 44 & 43.2 & & 41 & 52 & 12 &  2\,872 \\
WG600313P     & 09.-09. Jan. & 1993  &  0 & 44 & 50.4 & & 42 & 10 & 48 &  2\,672 \\
WG600368P     & 03.-03. Jan. & 1993  &  0 & 44 & 57.5 & & 41 & 21 & 00 &  2\,912 \\
WG600350P     & 05.-05. Jan. & 1993  &  0 & 45 &  2.4 & & 41 & 39 & 36 &  2\,680 \\
WG600332P     & 11.-11. Jan. & 1993  &  0 & 45 &  9.6 & & 41 & 58 & 12 &  2\,944 \\
WG600314P     & 10.-10. Jan. & 1993  &  0 & 45 & 14.3 & & 42 & 16 & 48 &  2\,704 \\
WG600369P     & 09.-09. Jan. & 1993  &  0 & 45 & 21.6 & & 41 & 27 &  0 &  2\,520 \\
WG600351P     & 02.-02. Jan. & 1993  &  0 & 45 & 28.7 & & 41 & 46 & 12 &  2\,824 \\
WG600333P     & 11.-11. Jan. & 1993  &  0 & 45 & 33.5 & & 42 &  4 & 48 &  3\,040 \\
WG600315P     & 06.-10. Jan. & 1993  &  0 & 45 & 40.7 & & 42 & 23 & 24 &  3\,024 \\
WG600370P     & 06.-06. Jan. & 1993  &  0 & 45 & 48.0 & & 41 & 33 & 36 &  2\,864 \\
WG600352P     & 10.-10. Jan. & 1993  &  0 & 45 & 52.7 & & 41 & 52 & 12 &  2\,632 \\
WG600334P     & 09.-09. Aug. & 1992  &  0 & 46 &  0.0 & & 42 & 10 & 48 &   720 \\
WG600334P-1   & 07.-07. Jan. & 1993  &  0 & 46 &  0.0 & & 42 & 10 & 48 &  2\,168 \\
WG600371P     & 08.-09. Aug. & 1992  &  0 & 46 & 12.0 & & 41 & 40 & 12 &  2\,080 \\
WG600353P     & 11.-11. Jan. & 1993  &  0 & 46 & 19.2 & & 41 & 58 & 48 &  3\,008 \\
WG600335P     & 09.-09. Aug. & 1992  &  0 & 46 & 24.0 & & 42 & 17 & 24 &   728 \\
WG600335P-1   & 10.-10. Jan. & 1993  &  0 & 46 & 24.0 & & 42 & 17 & 24 &  2\,368 \\
WG600372P     & 03.-03. July & 1993  &  0 & 46 & 38.4 & & 41 & 46 & 12 &  3\,088 \\
WG600354P     & 08.-08. Aug. & 1992  &  0 & 46 & 43.2 & & 42 &  4 & 48 &  2\,368 \\
WG600373P     & 11.-12. Jan. & 1993  &  0 & 47 &  2.4 & & 41 & 52 & 48 &  3\,144 \\
WG600355P     & 09.-09. Aug. & 1992  &  0 & 47 &  9.6 & & 42 & 11 & 24 &  1\,920 \\
WG600374P     & 09.-09. Aug. & 1992  &  0 & 47 & 26.3 & & 41 & 58 & 48 &   680 \\
WG600374P-1   & 04.-04. Jan. & 1993  &  0 & 47 & 26.3 & & 41 & 58 & 48 &  2\,008 \\
WG600375P     & 09.-09. Aug. & 1992  &  0 & 47 & 52.7 & & 42 &  5 & 24 &  1\,080 \\
WG600375P-1   & 15.-15. Jan. & 1993  &  0 & 47 & 52.7 & & 42 &  5 & 24 &  2\,112 \\
\end{supertabular}
\newline $^1$The coordinates give the centre of the FOV (nominal pointing direction).
\end{flushleft}

\clearpage
\scriptsize
\begin{table*}[h]
        \caption{\label{sourcelist92} \small List of all X-ray sources in M31 detected in the
              second ROSAT PSPC survey (SII).
              The meaning of the different columns
              is described in Sect. \ref{sourcedescript}. The listed
              count rate errors are only statistical. The systematic
              errors are expected to be less than $\sim15\%$.
              For sources not detected in a considered energy band
              $1\sigma$ upper limits have been calculated indicated by a
              `$<$`-symbol in front of the upper limit value.
              A conversion of count rates into fluxes
              depends on the assumed spectral shape. For M31-sources a power law
              with $\Gamma = -2.0$ and $N_{\rm H} = 9 \times 10^{20} \mbox{ cm}^{-2}$
              may be used, leading to the conversion factor
              $1 \mbox{ cts ksec}^{-1} = 3.00 \times 10^{-14} \mbox{ erg cm}^{-2} \mbox{ sec}^{-1}$
              in the 0.1 - 2.0 keV band ($B$-band). For foreground stars the application
              of this conversion factor leads to an over-estimate of the fluxes.}
\end{table*}
\tablehead{
\multicolumn{15}{l}{$\star$ Bulge sources}\\
\hline
  \multicolumn{1}{|c|}{SII}  & \multicolumn{2}{c}{R.A.} & \multicolumn{2}{c}{(J2000)} & \multicolumn{2}{c|}{Dec.} & $\sigma_{\rm Pos}$ & Cl. & 
     \multicolumn{1}{c|}{Maxlik} & \multicolumn{1}{c}{Rate ($B$)} & \multicolumn{1}{c}{Rate ($S$)} & \multicolumn{1}{c}{Rate ($H$)} & \multicolumn{1}{c}{Rate ($H_{\rm 1}$)} & 
     \multicolumn{1}{c|}{Rate ($H_{\rm 2}$)} \\
  \multicolumn{1}{|c|}{No.}  & (h) & (m) & (s) & ($\degr$) & ($'$) & ($''$) & ($''$) & & 
     \multicolumn{1}{c|}{(LH)} & \multicolumn{1}{c}{$(ct \cdot ks^{-1})$} & \multicolumn{1}{c}{$(ct \cdot ks^{-1})$} & \multicolumn{1}{c}{$(ct \cdot ks^{-1})$} & 
     \multicolumn{1}{c}{$(ct \cdot ks^{-1})$} & \multicolumn{1}{c|}{$(ct \cdot ks^{-1})$} \\
  \multicolumn{1}{|c|}{(1)}  & (2) & (3) & (4) & (5) & (6) & (7) & (8) & (9) & \multicolumn{1}{c|}{(10)} & \multicolumn{1}{c}{(11)} &
     \multicolumn{1}{c}{(12)} & \multicolumn{1}{c}{(13)} & \multicolumn{1}{c}{(14)} & \multicolumn{1}{c|}{(15)} \\ \hline \hline}
\tabletail{\hline}
\begin{flushleft}
\begin{supertabular}{|r\hs |r\hs r\hs r\Hs r\hs r\hs r\hs |r\hs r\hs |r\hs |r\hs r\hs r\hs r\hs r\hs|}
	  1 &  0 & 36 & 21.7 & 40 & 53 & 37 & 13  & 4 &   69.7 & $ <  0.61 \hspace{8ex}$ & $ <  0.20 \hspace{8ex}$ & $    5.73 \pm   0.65$ & $ <  1.70 \hspace{8ex}$ & $    4.22 \pm   0.53$ \\ 
	  2 &  0 & 36 & 43.0 & 41 &  9 &  2 & 19  & 4 &   27.1 & $ <  0.43 \hspace{8ex}$ & $ <  0.16 \hspace{8ex}$ & $    3.18 \pm   0.71$ & $ <  0.87 \hspace{8ex}$ & $    3.56 \pm   0.61$ \\ 
	  3 &  0 & 36 & 49.6 & 40 & 55 & 12 & 18  & 4 &   11.8 & $ <  0.13 \hspace{8ex}$ & $ <  0.14 \hspace{8ex}$ & $    1.72 \pm   0.42$ & $ <  0.99 \hspace{8ex}$ & $ <  1.20 \hspace{8ex}$ \\ 
	  4 &  0 & 36 & 49.8 & 40 &  8 & 52 & 12  & 1 &   45.0 & $   10.62 \pm   1.82$ & $ <  3.50 \hspace{8ex}$ & $    7.46 \pm   1.43$ & $ <  3.59 \hspace{8ex}$ & $    3.74 \pm   1.04$ \\ 
	  5 &  0 & 36 & 57.4 & 40 & 26 & 10 & 12  & 1 &   12.2 & $ <  2.13 \hspace{8ex}$ & $ <  0.62 \hspace{8ex}$ & $    1.34 \pm   0.43$ & $ <  0.81 \hspace{8ex}$ & $ <  1.30 \hspace{8ex}$ \\ 
	  6 &  0 & 37 & 11.4 & 40 & 22 & 10 & 11  & 1 &   10.5 & $ <  1.65 \hspace{8ex}$ & $ <  0.70 \hspace{8ex}$ & $ <  1.13 \hspace{8ex}$ & $ <  0.63 \hspace{8ex}$ & $ <  0.55 \hspace{8ex}$ \\ 
	  7 &  0 & 37 & 18.7 & 40 & 29 & 47 &  8  & 1 &   82.4 & $    3.43 \pm   0.65$ & $ <  0.30 \hspace{8ex}$ & $    3.90 \pm   0.60$ & $    2.02 \pm   0.43$ & $    1.79 \pm   0.41$ \\ 
	  8 &  0 & 37 & 22.5 & 40 & 43 & 49 &  9  & 1 &   86.3 & $   10.19 \pm   1.34$ & $    5.09 \pm   0.98$ & $    4.93 \pm   0.90$ & $    2.97 \pm   0.69$ & $    1.82 \pm   0.56$ \\
\end{supertabular}\newline
\end{flushleft}
\normalsize

\begin{landscape}

\clearpage
\scriptsize
\begin{table*}[h]
        \caption{\label{sourcelist} \small Total list of all ROSAT PSPC X-ray 
              sources in \object{M31} merged from the source lists of both surveys.
              The different symbols in front of the
              RXJ numbers in column (1) are explained at the top of the
              table. The meaning of the different columns
              is described in Sect. \ref{sourcedescript}. The listed
              count rate errors are purely statistical. The systematic
              errors are expected to be less than $\sim15\%$.
              Count rates for bulge sources (marked with $\star$) may
              be uncertain due to confusion.
              For sources not detected in a considered energy band, 
              $1\sigma$ upper limits have been calculated indicated by a
              `$<$`-symbol in front of the upper limit value.
              A conversion of count rates into fluxes
              depends on the assumed spectral shape. For \object{M31}-sources a power law
              with photon index $\Gamma = -2.0$ and $N_{\rm H} = 9 \times 10^{20} \mbox{ cm}^{-2}$
              may be used, leading to a conversion factor
              $1 \mbox{ ct ks}^{-1} = 3.00 \times 10^{-14} \mbox{ erg cm}^{-2} \mbox{ s}^{-1}$
              in the 0.1 - 2.0 keV band ($B$-band). For foreground stars, the application
              of this conversion factor leads to an overestimation of the flux.}
\end{table*}
\tablehead{
\multicolumn{16}{l}{$^a$ Foreground star \hspace{3ex} $^b$ Galaxy \hspace{3ex} $^c$ Supersoft source candidate \hspace{3ex} $^d$ Globular cluster \hspace{3ex} $^e$ SNR}\\
\multicolumn{16}{l}{$\star$ bulge source \hspace{3ex} $\sim$ Variable source (ref. Tables \ref{varEinsteinList}, \ref{varSurveysList} and footnote in Sect.\ \ref{corSurveyI}) \hspace{3ex} $\dagger$ Source with uncertain count rate.}\\
\hline
  \multicolumn{1}{|c}{RXJ}  & SI & & \multicolumn{2}{c}{R.A.} & \multicolumn{2}{c}{(J2000)} & \multicolumn{2}{c|}{Dec.} & $\sigma_{\rm Pos}$ & Cl. & 
     \multicolumn{1}{c}{Maxlik} & \multicolumn{1}{c}{Rate ($B$)} & \multicolumn{1}{c}{Rate ($S$)} & \multicolumn{1}{c}{Rate ($H$)} & \multicolumn{1}{c}{Rate ($H_{\rm 1}$)} & \multicolumn{1}{c|}{Rate ($H_{\rm 2}$)} \\
  \multicolumn{1}{|c}{No.}  & No. & & (h) & (m) & (s) & ($\degr$) & ($'$) & ($''$) & ($''$) & & 
     \multicolumn{1}{c}{(LH)} & \multicolumn{1}{c}{$(ct \cdot ks^{-1})$} & \multicolumn{1}{c}{$(ct \cdot ks^{-1})$} & \multicolumn{1}{c}{$(ct \cdot ks^{-1})$} & 
     \multicolumn{1}{c}{$(ct \cdot ks^{-1})$} & \multicolumn{1}{c|}{$(ct \cdot ks^{-1})$} \\
  \multicolumn{1}{|c}{(1)}  & (2) & & (3) & (4) & (5) & (6) & (7) & (8) & (9) & (10) & \multicolumn{1}{c}{(11)} & \multicolumn{1}{c}{(12)} &
     \multicolumn{1}{c}{(13)} & \multicolumn{1}{c}{(14)} & \multicolumn{1}{c}{(15)} & \multicolumn{1}{c|}{(16)} \\ \hline \hline}
\tabletail{\hline}
\begin{flushleft}
\begin{supertabular}{|r\hs r\ns l\hs |r\hs r\hs r\Hs r\hs r\hs r\hs |r\hs r\hs |r\hs r\hs r\hs r\hs r\hs r\hs|}
$^d$RX J0036.3+4053 &    & &  0 & 36 & 21.7 & 40 & 53 & 37 & 13  & 4 &   69.7 & $ <  0.61 \hspace{8ex}$ & $ <  0.20 \hspace{8ex}$ & $    5.73 \pm   0.65$ & $ <  1.70 \hspace{8ex}$ & $    4.22 \pm   0.53$ \\ 
RX J0036.7+4109 &    & &  0 & 36 & 43.0 & 41 &  9 &  2 & 19  & 4 &   27.1 & $ <  0.43 \hspace{8ex}$ & $ <  0.16 \hspace{8ex}$ & $    3.18 \pm   0.71$ & $ <  0.87 \hspace{8ex}$ & $    3.56 \pm   0.61$ \\ 
RX J0036.8+4008 &    & &  0 & 36 & 49.8 & 40 &  8 & 52 & 12  & 1 &   45.0 & $   10.62 \pm   1.82$ & $ <  3.50 \hspace{8ex}$ & $    7.46 \pm   1.43$ & $ <  3.59 \hspace{8ex}$ & $    3.74 \pm   1.04$ \\ 
RX J0036.8+4055 &    & &  0 & 36 & 49.6 & 40 & 55 & 12 & 18  & 4 &   11.8 & $ <  0.13 \hspace{8ex}$ & $ <  0.14 \hspace{8ex}$ & $    1.72 \pm   0.42$ & $ <  0.99 \hspace{8ex}$ & $ <  1.20 \hspace{8ex}$ \\ 
RX J0036.9+4026 &    & &  0 & 36 & 57.4 & 40 & 26 & 10 & 12  & 1 &   12.2 & $ <  2.13 \hspace{8ex}$ & $ <  0.62 \hspace{8ex}$ & $    1.34 \pm   0.43$ & $ <  0.81 \hspace{8ex}$ & $ <  1.30 \hspace{8ex}$ \\ 
RX J0037.1+4022 &    & &  0 & 37 & 11.4 & 40 & 22 & 10 & 11  & 1 &   10.5 & $ <  1.65 \hspace{8ex}$ & $ <  0.70 \hspace{8ex}$ & $ <  1.13 \hspace{8ex}$ & $ <  0.63 \hspace{8ex}$ & $ <  0.55 \hspace{8ex}$ \\ 
RX J0037.2+4045 &   1& &  0 & 37 & 17.4 & 40 & 45 & 23 &  6  & 2 &   26.7 & $    2.14 \pm   0.80$ & $    0.95 \pm   0.80$ & $ <  0.87 \hspace{8ex}$ & $ <  0.12 \hspace{8ex}$ & $ <  0.80 \hspace{8ex}$ \\ 
RX J0037.3+4029 &   2&-&  0 & 37 & 18.7 & 40 & 29 & 47 &  8  & 1 &   82.4 & $    3.43 \pm   0.65$ & $ <  0.30 \hspace{8ex}$ & $    3.90 \pm   0.60$ & $    2.02 \pm   0.43$ & $    1.79 \pm   0.41$ \\
\end{supertabular}\newline
\end{flushleft}
\normalsize

\end{landscape}

\begin{table*}[h]
	\caption{\label{ID-Einstein} \small Table of the identifications of
                  ROSAT PSPC sources with {\it Einstein} sources listed
                  by TF. $F_{\rm R}$ gives the ROSAT
                  source flux using the {\it Einstein} spectral model of TF and
                  $F_{\rm E}$ gives the {\it Einstein} source flux of the correlated
                  {\it Einstein} source (see Sect.\ \ref{corein}). 
                  The distance between two correlating
                  sources is given in arcseconds $(\arcsec$) as well as in
                  units of the combined positional error ($\sigma$) of both
                  sources. The last column gives the flux ratio between 
                  the ROSAT and the {\it Einstein} measurements, showing 
                  possible long term variabilities between the epochs of
                  the two observations. Sources with ROSAT numbers preceeded by a
                  $\star$ belong to the bulge region.} 
\end{table*}
\tablehead{\hline
\multicolumn{1}{c}{ROSAT}  & \multicolumn{1}{c}{$F_{\rm R} \, (\times 10^{13})$} &
   {\it Einstein} & \multicolumn{1}{c}{$F_{\rm E} \, (\times 10^{13})$} & \multicolumn{2}{c}{Distance} &
   \multicolumn{1}{c}{$F_{\rm R} / F_{\rm E}$}\\
\multicolumn{1}{c}{No.} & \multicolumn{1}{c}{(cgs)} & \multicolumn{1}{c}{No.} & \multicolumn{1}{c}{(cgs)} &
   $('')$ & $(\sigma)$ \\ \hline}
\tabletail{\hline}
\begin{supertabular}{rrrrrrr}
			RX J0039.4+4035  & $   0.17 \pm  0.06 $ &   1 & $   0.60 \pm  0.26 $ &  59.1 & 1.31 & $  0.28 \pm 0.16 $ \\
			RX J0040.0+4031  & $   1.04 \pm  0.10 $ &   2 & $   1.64 \pm  0.42 $ &  31.7 & 0.70 & $  0.63 \pm 0.17 $ \\
			RX J0040.2+4050  & $  31.64 \pm  0.48 $ &   3 & $  25.38 \pm  2.50 $ &   3.6 & 0.59 & $  1.25 \pm 0.12 $ \\
			RX J0040.3+4043  & $  13.82 \pm  0.29 $ &   4 & $  16.35 \pm  2.09 $ &   3.4 & 0.58 & $  0.85 \pm 0.11 $ \\
			RX J0040.4+4029  & $   1.07 \pm  0.10 $ &   5 & $   0.99 \pm  0.28 $ &  13.4 & 0.30 & $  1.08 \pm 0.33 $ \\
			RX J0040.4+4129  & $   2.30 \pm  0.41 $ &   6 & $   1.06 \pm  0.35 $ &   9.5 & 0.20 & $  2.17 \pm 0.81 $ \\
			RX J0040.7+4051  & $   0.49 \pm  0.07 $ &   7 & $   0.66 \pm  0.26 $ &   6.0 & 0.13 & $  0.74 \pm 0.30 $ \\
			RX J0041.4+4058  & $   1.94 \pm  0.12 $ &   8 & $   1.77 \pm  0.50 $ &  24.6 & 0.54 & $  1.09 \pm 0.32 $ \\
			RX J0041.7+4134  & $  14.38 \pm  0.49 $ &   9 & $   8.72 \pm  1.08 $ &   1.8 & 0.29 & $  1.65 \pm 0.21 $ \\
			RX J0041.8+4021  & $  24.25 \pm  0.76 $ &  11 & $  15.54 \pm  0.88 $ &   5.4 & 0.12 & $  1.56 \pm 0.10 $ \\
			RX J0041.8+4113  & $   0.43 \pm  0.09 $ &  10 & $   0.90 \pm  0.24 $ &  32.9 & 0.73 & $  0.48 \pm 0.16 $ \\
			RX J0042.2+4019  & $  40.38 \pm  1.23 $ &  15 & $  48.83 \pm  1.61 $ &   2.9 & 0.06 & $  0.83 \pm 0.04 $ \\
			RX J0042.2+4039  & $   2.18 \pm  0.19 $ &  13 & $   1.64 \pm  0.38 $ &  18.4 & 0.41 & $  1.33 \pm 0.33 $ \\
			RX J0042.2+4055  & $   2.71 \pm  0.18 $ &  18 & $   1.67 \pm  0.48 $ &   1.9 & 0.32 & $  1.62 \pm 0.48 $ \\
			RX J0042.2+4101  & $   9.53 \pm  0.32 $ &  16 & $   3.88 \pm  0.75 $ &   0.7 & 0.12 & $  2.46 \pm 0.48 $ \\
			RX J0042.2+4112  & $   9.04 \pm  0.25 $ &  19 & $   4.26 \pm  0.54 $ &   8.0 & 1.37 & $  2.12 \pm 0.28 $ \\
			RX J0042.2+4117  & $   2.18 \pm  0.16 $ &  17 & $   1.26 \pm  0.38 $ &   5.0 & 0.86 & $  1.73 \pm 0.54 $ \\
			RX J0042.2+4118  & $   9.71 \pm  0.32 $ &  14 & $   3.23 \pm  0.51 $ &   9.5 & 1.63 & $  3.01 \pm 0.49 $ \\
  $\star$RX J0042.3+4113  & $   5.02 \pm  0.23 $ &  20 & $   4.93 \pm  0.54 $ &   9.1 & 1.55 & $  1.02 \pm 0.12 $ \\
  $\star$RX J0042.3+4115  & $  16.06 \pm  0.41 $ &  23 & $   6.88 \pm  0.61 $ &   1.0 & 0.17 & $  2.33 \pm 0.21 $ \\
			RX J0042.4+4104  & $   4.78 \pm  0.22 $ &  28 & $   3.07 \pm  0.71 $ &   6.4 & 0.99 & $  1.56 \pm 0.37 $ \\
			RX J0042.4+4112  & $   5.92 \pm  0.25 $ &  27 & $   3.38 \pm  0.52 $ &   5.2 & 0.89 & $  1.75 \pm 0.28 $ \\
			RX J0042.4+4125  & $   1.13 \pm  0.13 $ &  30 & $   1.71 \pm  0.47 $ &  38.7 & 0.85 & $  0.66 \pm 0.20 $ \\
			RX J0042.5+4103  & $   1.34 \pm  0.13 $ &  37 & $   2.61 \pm  0.85 $ &  65.8 & 1.45 & $  0.51 \pm 0.17 $ \\
  $\star$RX J0042.5+4113  & $   4.15 \pm  0.22 $ &  34 & $   3.05 \pm  0.50 $ &   7.1 & 1.22 & $  1.36 \pm 0.23 $ \\
  $\star$RX J0042.5+4116  & $   6.01 \pm  0.21 $ &  32 & $   3.62 \pm  0.51 $ &   6.8 & 1.16 & $  1.66 \pm 0.24 $ \\
  $\star$RX J0042.5+4119  & $   3.30 \pm  0.20 $ &  33 & $   0.83 \pm  0.28 $ &   9.2 & 1.58 & $  3.97 \pm 1.35 $ \\
			RX J0042.5+4132  & $   1.69 \pm  0.16 $ &  38 & $   1.56 \pm  0.29 $ &   5.0 & 0.11 & $  1.08 \pm 0.23 $ \\
			RX J0042.6+4052  & $  32.45 \pm  0.53 $ &  51 & $   9.16 \pm  1.01 $ &  11.6 & 1.98 & $  3.54 \pm 0.40 $ \\
  $\star$RX J0042.6+4114  & $   3.75 \pm  0.17 $ &  48 & $   0.97 \pm  0.27 $ &   9.1 & 1.56 & $  3.87 \pm 1.09 $ \\
  $\star$RX J0042.6+4115  & $  42.30 \pm  0.38 $ &  41 & $  41.28 \pm  1.22 $ &   6.9 & 1.19 & $  1.02 \pm 0.03 $ \\
  $\star$RX J0042.7+4111  & $   4.13 \pm  0.21 $ &  58 & $   1.59 \pm  0.44 $ &   3.7 & 0.63 & $  2.60 \pm 0.73 $ \\
  $\star$RX J0042.7+4115  & $  22.80 \pm  0.30 $ &  52 & $   1.62 \pm  0.31 $ &   4.5 & 0.77 & $ 14.07 \pm 2.71 $ \\
  $\star$RX J0042.7+4116  & $  29.89 \pm  0.28 $ &  59 & $   1.81 \pm  0.34 $ &   8.6 & 1.47 & $ 16.51 \pm 3.14 $ \\
  $\star$RX J0042.8+4115  & $  17.39 \pm  0.25 $ &  63 & $   8.04 \pm  0.63 $ &   5.6 & 0.96 & $  2.16 \pm 0.17 $ \\
  $\star$RX J0042.8+4118  & $   9.68 \pm  0.33 $ &  68 & $   5.64 \pm  0.58 $ &   6.4 & 1.09 & $  1.72 \pm 0.19 $ \\
			RX J0042.8+4125  & $   3.95 \pm  0.22 $ &  62 & $   3.44 \pm  0.84 $ &   6.2 & 1.06 & $  1.15 \pm 0.29 $ \\
			RX J0042.8+4131  & $  18.62 \pm  0.45 $ &  67 & $  11.95 \pm  1.10 $ &   4.6 & 0.78 & $  1.56 \pm 0.15 $ \\
			RX J0042.9+4111  & $   3.84 \pm  0.21 $ &  73 & $   3.13 \pm  0.51 $ &   5.4 & 0.92 & $  1.23 \pm 0.21 $ \\
  $\star$RX J0042.9+4117  & $   8.46 \pm  0.13 $ &  72 & $   0.77 \pm  0.27 $ &  10.6 & 1.83 & $ 10.99 \pm 3.82 $ \\
  $\star$RX J0042.9+4119  & $   3.66 \pm  0.22 $ &  76 & $   1.64 \pm  0.46 $ &   4.6 & 0.78 & $  2.23 \pm 0.64 $ \\
			RX J0042.9+4125  & $   3.05 \pm  0.19 $ &  70 & $   3.77 \pm  0.34 $ &   2.3 & 0.05 & $  0.81 \pm 0.09 $ \\
  $\star$RX J0043.0+4115  & $   8.97 \pm  0.31 $ &  79 & $   3.26 \pm  0.42 $ &   1.7 & 0.30 & $  2.75 \pm 0.36 $ \\
  $\star$RX J0043.0+4117  & $   6.19 \pm  0.27 $ &  80 & $   1.92 \pm  0.35 $ &   9.1 & 1.56 & $  3.22 \pm 0.61 $ \\
			RX J0043.1+4048  & $   1.48 \pm  0.17 $ &  81 & $   1.04 \pm  0.42 $ &  52.4 & 1.15 & $  1.43 \pm 0.60 $ \\
			RX J0043.1+4114  & $   8.60 \pm  0.30 $ &  83 & $   7.51 \pm  0.61 $ &   5.8 & 1.00 & $  1.15 \pm 0.10 $ \\
			RX J0043.1+4118  & $   7.54 \pm  0.24 $ &  82 & $   2.03 \pm  0.31 $ &   2.4 & 0.41 & $  3.72 \pm 0.58 $ \\
			RX J0043.2+4107  & $   3.08 \pm  0.19 $ &  85 & $   4.27 \pm  0.94 $ &   5.4 & 0.93 & $  0.72 \pm 0.16 $ \\
			RX J0043.3+4117  & $   4.44 \pm  0.25 $ &  88 & $   1.31 \pm  0.34 $ &  32.7 & 0.72 & $  3.39 \pm 0.89 $ \\
			RX J0043.3+4120  & $   1.80 \pm  0.17 $ &  87 & $   2.15 \pm  0.52 $ &   5.0 & 0.85 & $  0.84 \pm 0.22 $ \\
			RX J0043.3+4159  & $   1.91 \pm  0.22 $ &  86 & $   0.96 \pm  0.29 $ &  12.1 & 0.27 & $  1.99 \pm 0.64 $ \\
			RX J0043.4+4107  & $   1.67 \pm  0.15 $ &  90 & $   1.64 \pm  0.28 $ &  15.3 & 0.34 & $  1.02 \pm 0.20 $ \\
			RX J0043.4+4118  & $   3.20 \pm  0.20 $ &  89 & $   2.48 \pm  0.60 $ &  15.5 & 1.99 & $  1.29 \pm 0.32 $ \\
			RX J0043.5+4113  & $   6.31 \pm  0.27 $ &  91 & $   6.24 \pm  1.02 $ &   8.8 & 1.51 & $  1.01 \pm 0.17 $ \\
			RX J0043.6+4114  & $   5.13 \pm  0.23 $ &  92 & $   4.73 \pm  0.90 $ &   9.3 & 1.41 & $  1.08 \pm 0.21 $ \\
			RX J0043.7+4124  & $   1.00 \pm  0.13 $ &  93 & $   2.86 \pm  0.74 $ &   6.4 & 0.68 & $  0.35 \pm 0.10 $ \\
			RX J0043.8+4116  & $   6.29 \pm  0.26 $ &  94 & $   4.05 \pm  0.78 $ &   5.3 & 0.82 & $  1.55 \pm 0.30 $ \\
			RX J0043.9+4045  & $   0.85 \pm  0.19 $ &  95 & $   1.64 \pm  0.50 $ &  61.8 & 1.34 & $  0.52 \pm 0.19 $ \\
			RX J0044.4+4121  & $   7.97 \pm  0.30 $ &  97 & $   8.59 \pm  1.09 $ &   2.6 & 0.45 & $  0.93 \pm 0.12 $ \\
			RX J0044.9+4059  & $   0.84 \pm  0.17 $ &  98 & $   2.07 \pm  0.73 $ &  47.2 & 1.03 & $  0.40 \pm 0.16 $ \\
			RX J0045.2+4136  & $   0.40 \pm  0.08 $ &  99 & $   1.03 \pm  0.27 $ &  52.7 & 1.15 & $  0.39 \pm 0.13 $ \\
			RX J0045.4+4146  & $   0.32 \pm  0.32 $ & 100 & $   0.71 \pm  0.23 $ &  70.4 & 1.53 & $  0.46 \pm 0.48 $ \\
			RX J0045.6+4208  & $   6.92 \pm  0.30 $ & 101 & $   4.99 \pm  0.86 $ &   0.6 & 0.09 & $  1.39 \pm 0.25 $ \\
			RX J0045.7+4139  & $  36.00 \pm  0.52 $ & 102 & $  43.18 \pm  3.15 $ &   8.3 & 1.37 & $  0.83 \pm 0.06 $ \\
			RX J0046.1+4208  & $   1.03 \pm  0.12 $ & 103 & $   0.95 \pm  0.24 $ &  28.2 & 0.62 & $  1.08 \pm 0.30 $ \\
			RX J0046.4+4201  & $  10.08 \pm  0.31 $ & 105 & $   5.52 \pm  0.89 $ &   2.1 & 0.34 & $  1.83 \pm 0.30 $ \\
			RX J0046.4+4204  & $   8.88 \pm  0.31 $ & 104 & $   8.56 \pm  0.48 $ &  15.1 & 0.33 & $  1.04 \pm 0.07 $ \\
			RX J0046.9+4220  & $   5.19 \pm  0.24 $ & 107 & $   3.54 \pm  0.53 $ &  55.6 & 1.23 & $  1.47 \pm 0.23 $ \\
			RX J0048.0+4140  & $   7.43 \pm  0.55 $ & 108 & $   4.46 \pm  0.89 $ &  44.1 & 0.97 & $  1.67 \pm 0.35 $ \\
\end{supertabular}
\normalsize

\small
\begin{table*}[h]
	\caption{\label{ID-Tabelle} \small Table of all optical and radio
                 identifications.}
\end{table*}
\tablehead{\hline
ROSAT  & Obj.-Cl. & Identification & \multicolumn{2}{c}{Distance} \\
No.    &          &                & $('')$ & $(\sigma)$ \\ \hline}
\tabletail{\hline}
\begin{supertabular}{cllrr}
RX J0036.3+4053  & GC                   & BA87(5)                                                      & 10.4 & 0.78 \\
RX J0037.3+4043  & Star  (F5)           & SIMBAD(\object{SAO 36516})                                   & 13.0 & 1.49 \\
RX J0038.0+4026  & Star                 & HA94(38232)                                                  &  6.1 & 1.21 \\
RX J0038.4+4012  & Star                 & HA94(9276)                                                   &  6.9 & 1.38 \\
RX J0038.4+4136  & EO                   & \object{87GB 003540.8+412038}                                & 34.9 & 1.51 \\
RX J0038.6+4026  & Star  (K0)           & SIMBAD(\object{SAO 36541})                                   &  0.8 & 0.16 \\
RX J0039.5+4008  & EO                   & \object{B3 0036+398}                                         & 10.2 & 1.83 \\
RX J0039.6+4011  & Star                 & HA94(7297)                                                   &  4.1 & 0.47 \\
RX J0039.7+4039  & Star                 & HA94(81094)                                                  & 11.0 & 1.25 \\
RX J0040.1+4006  & Star                 & HA94(2646)                                                   & 15.9 & 1.17 \\
RX J0040.1+4044  & Star                 & HA94(101871)                                                 & 12.7 & 1.47 \\
RX J0040.1+4047  & Star                 & HA94(111215)                                                 &  7.0 & 1.40 \\
RX J0040.2+4015  & Star                 & HA94(13652)                                                  &  2.4 & 0.48 \\
RX J0040.2+4050  & EO                   & \object{87GB 003730.5+403346}                                &  6.0 & 1.11 \\
RX J0040.3+4043  & GC                   & MA94a(6)                                                     &  1.1 & 0.22 \\
RX J0040.4+4050  & Star                 & HA94(119503)                                                 &  4.2 & 0.85 \\
RX J0040.4+4129  & GC                   & BA87(51)                                                     & 20.4 & 1.76 \\
RX J0040.5+4033  & GC                   & MA94a(16)                                                    &  3.8 & 0.24 \\
RX J0040.5+4034  & Star                 & HA94(61478)                                                  &  4.0 & 0.56 \\
RX J0040.7+4055  & SNR                  & DO80(7)                                                      & 11.3 & 0.64 \\
RX J0040.8+4011  & Star  (K2V)          & SIMBAD(\object{GJ 28})                                       &  5.4 & 1.04 \\
RX J0040.9+4056  & Star                 & HA94(140092)                                                 &  4.7 & 0.94 \\
RX J0041.3+4012  & EO                   & \object{87GB 003844.0+395608}                                &  4.8 & 0.26 \\
RX J0041.3+4051  & Star                 & HA94(125532)                                                 &  3.6 & 0.71 \\
RX J0041.3+4109  & Star                 & HA94(213303)                                                 & 56.0 & 1.86 \\
RX J0041.4+4025  & Star                 & HA94(37411)                                                  & 14.9 & 1.63 \\
RX J0041.5+4106  & SNR                  & MA95(3-041),DO80(11)                                         &  3.2 & 0.45 \\
RX J0041.6+4103  & Star                 & HA94(175678)                                                 & 14.9 & 1.90 \\
RX J0041.6+4112  & Star                 & HA94(232216)                                                 &  8.1 & 1.61 \\
RX J0041.7+4105  & Star                 & HA94(189885)                                                 &  0.4 & 0.08 \\
RX J0041.7+4134  & GC                   & BA87(98)                                                     &  1.6 & 0.28 \\
RX J0041.8+4021  & EO                   & \object{MRK 0957}                                            & 11.0 & 1.93 \\
RX J0041.9+4046  & SNR                  & MA95(2-021)                                                  & 26.0 & 1.67 \\
RX J0042.0+4031  & Star                 & HA94(50780)                                                  &  8.0 & 0.82 \\
RX J0042.0+4033  & Star                 & HA94(56227)                                                  & 17.3 & 1.78 \\
RX J0042.0+4041  & Star  (F5)           & SIMBAD(\object{HD 3914})                                     &  5.5 & 1.05 \\
RX J0042.0+4102  & GC                   & MA94a(130)                                                   &  9.0 & 1.76 \\
RX J0042.1+4016  & Star                 & HA94(16045)                                                  & 10.1 & 1.10 \\
RX J0042.2+4101  & GC                   & BA87(138),MA94a(159)                                         &  2.0 & 0.40 \\
RX J0042.2+4105  & Star                 & HA94(196074)                                                 &  4.1 & 0.81 \\
RX J0042.2+4118  & Star                 & HA94(266451)                                                 &  6.4 & 1.29 \\
RX J0042.3+4113  & GC                   & BA87(142),MA94a(164)                                         &  7.3 & 1.43 \\
RX J0042.3+4126  & Star                 & HA94(302803)                                                 & 10.5 & 1.21 \\
RX J0042.3+4129  & EO                   & \object{87GB 003934.6+411250}                                & 11.9 & 1.55 \\
RX J0042.4+4055  & Star                 & HA94(136397)                                                 &  8.7 & 1.74 \\
RX J0042.4+4057  & GC                   & BA87(153),MA94a(173)                                         &  6.2 & 1.22 \\
RX J0042.4+4108  & Star                 & HA94(208922)                                                 & 10.7 & 1.37 \\
RX J0042.4+4129  & Star                 & HA94(316946)                                                 &  2.4 & 0.32 \\
RX J0042.5+4103  & GC                   & BA87(171),MA94a(196)                                         &  2.7 & 0.53 \\
RX J0042.5+4119  & GC                   & BA87(168),BA93(1),MA94a(192)                                 &  4.6 & 0.91 \\
RX J0042.5+4132  & GC                   & BA87(176),MA94a(205)                                         &  2.5 & 0.39 \\
RX J0042.6+4052  & EO                   & \object{NGC 0221} = \object{M32}                             &  9.7 & 1.90 \\
RX J0042.6+4115  & GC                   & BA93(21)                                                     &  7.0 & 1.38 \\
RX J0042.8+4125  & SNR                  & DO80(13)                                                     & 20.4 & 1.15 \\
RX J0042.8+4131  & GC                   & BA87(196),MA94a(225)                                         &  1.6 & 0.31 \\
RX J0042.9+4119  & GC                   & BA87(203)                                                    &  7.2 & 1.42 \\
RX J0042.9+4125  & SNR                  & DO80(13)                                                     & 14.4 & 0.79 \\
RX J0043.0+4110  & Star                 & HA94(223193)                                                 &  3.5 & 0.70 \\
RX J0043.0+4115  & GC                   & BA87(206)                                                    &  3.0 & 0.60 \\
RX J0043.0+4117  & GC                   & BA87(208)                                                    &  6.1 & 1.20 \\
RX J0043.0+4121  & GC                   & BA87(207),MA94a(240)                                         &  4.3 & 0.84 \\
RX J0043.0+4130  & GC                   & MA94a(236)                                                   &  4.1 & 0.80 \\
RX J0043.1+4114  & GC                   & BA87(214),MA94a(251)                                         &  3.0 & 0.60 \\
RX J0043.2+4107  & GC                   & BA87(220),MA94a(257)                                         &  5.5 & 1.08 \\
RX J0043.2+4112  & GC                   & BA87(226),MA94a(269)                                         & 11.5 & 0.94 \\
RX J0043.2+4127  & GC                   & BA87(225),MA94a(266)                                         &  6.0 & 1.02 \\
RX J0043.3+4114  & Star                 & HA94(239273)                                                 & 16.0 & 1.53 \\
RX J0043.4+4118  & SNR                  & MA95(2-032),DO80(15)                                         & 14.2 & 1.63 \\
RX J0043.5+4116  & Star                 & HA94(254312)                                                 &  1.1 & 0.09 \\
RX J0043.6+4054  & EO                   & \object{B3 0040+406}                                         & 11.5 & 1.31 \\
RX J0043.6+4114  & GC                   & BA87(246),MA94a(299)                                         &  7.1 & 1.18 \\
RX J0043.6+4126  & SNR                  & MA95(3-059),DO80(16)                                         &  9.2 & 1.02 \\
RX J0043.6+4138  & Star                 & HA94(375268)                                                 &  8.3 & 1.65 \\
RX J0043.7+4128  & GC                   & MA94a(311)                                                   &  2.1 & 0.26 \\
RX J0043.7+4136  & GC                   & MA94a(314)                                                   &  4.2 & 0.83 \\
RX J0043.8+4106  & Star                 & HA94(196928)                                                 & 16.2 & 1.78 \\
RX J0043.8+4111  & SNR                  & BW93(230A)                                                   & 10.7 & 0.85 \\
RX J0043.8+4127  & GC                   & MA94a(317)                                                   & 39.9 & 0.83 \\
RX J0043.9+4113  & SNR                  & MA95(2-038),DO80(18),BW93(252)                               & 24.7 & 1.78 \\
RX J0043.9+4122  & GC                   & BA87(267),MA94a(334)                                         &  3.0 & 0.58 \\
RX J0043.9+4127  & Star                 & HA94(307124)                                                 &  7.6 & 1.53 \\
RX J0043.9+4152  & SNR                  & MA95(2-037),DO80(17)                                         & 11.7 & 1.06 \\
RX J0043.9+4157  & EO                   & \object{87GB 004113.4+414049}                                &  7.4 & 0.87 \\
RX J0044.0+4149  & SNR                  & MA95(3-072)                                                  & 19.5 & 1.51 \\
RX J0044.2+4119  & SNR                  & MA95(3-079),BW93(327)                                        &  9.1 & 1.28 \\
RX J0044.2+4126  & Star                 & HA94(302441)                                                 &  7.5 & 0.75 \\
RX J0044.4+4121  & GC                   & MA94a(387)                                                   &  1.9 & 0.36 \\
RX J0044.4+4136  & GC                   & MA94a(380)                                                   &  5.8 & 1.14 \\
RX J0044.6+4125  & SNR                  & MA95(3-086),BW93(490A)                                       &  4.9 & 0.58 \\
RX J0044.8+4129  & SNR                  & MA95(B-012),BW93(566)                                        &  8.4 & 1.19 \\
RX J0044.8+4229  & Star  (G5)           & SIMBAD(\object{HD4194})                                      &  6.6 & 0.56 \\
RX J0045.1+4202  & Star                 & HA94(441416)                                                 &  6.4 & 1.28 \\
RX J0045.2+4136  & SNR                  & MA95(2-048),BW93(717),DO80(19)                               & 11.9 & 1.23 \\
RX J0045.2+4217  & Star                 & HA94(466304)                                                 & 14.8 & 1.48 \\
RX J0045.4+4132  & GC                   & MA94a(447)                                                   & 15.5 & 1.85 \\
RX J0045.4+4146  & SNR                  & MA95(1-013)                                                  & 17.3 & 1.54 \\
RX J0045.5+4210  & Star                 & HA94(456238)                                                 &  7.8 & 1.48 \\
RX J0045.7+4139  & GC                   & BA87(318),MA94a(468)                                         &  5.8 & 1.08 \\
RX J0045.9+4156  & Star                 & HA94(431022)                                                 &  6.1 & 1.21 \\
RX J0045.9+4203  & Star                 & HA94(442137)                                                 &  8.5 & 1.10 \\
RX J0045.9+4226  & Star                 & HA94(479836)                                                 &  8.1 & 1.61 \\
RX J0046.0+4136  & Star                 & HA94(364424)                                                 & 20.4 & 1.78 \\
RX J0046.2+4154  & Star                 & HA94(425498)                                                 &  6.3 & 1.26 \\
RX J0046.4+4201  & GC                   & BA87(329),MA94a(487)                                         &  3.0 & 0.53 \\
RX J0046.6+4225  & Star                 & HA94(478682)                                                 &  3.8 & 0.76 \\
RX J0046.7+4208  & EO                   & \object{87GB 004359.8+415217}                                &  6.1 & 1.20 \\
RX J0046.7+4230  & Star                 & HA94(483376)                                                 & 18.9 & 1.86 \\
RX J0047.0+4157  & Star  (F)            & SIMBAD(\object{HD4444})                                      &  1.9 & 0.22 \\
RX J0047.0+4201  & Star                 & HA94(440331)                                                 &  9.8 & 0.66 \\
RX J0047.2+4202  & Star                 & HA94(441062)                                                 & 14.5 & 1.92 \\
RX J0047.4+4220  & Star                 & HA94(472654)                                                 &  7.7 & 1.54 \\
RX J0047.4+4221  & Star                 & HA94(474007)                                                 &  5.6 & 1.12 \\
RX J0047.7+4201  & Star                 & HA94(439698)                                                 & 12.8 & 1.46 \\
RX J0048.4+4157  & Star  (F8)           & SIMBAD(\object{SAO 36677})                                   &  7.4 & 1.14 \\
\end{supertabular}
\normalsize



\begin{thebibliography}{}

\bibitem[1987]{Bat87}  Battistini P.L., B\`onoli F., Braccesi A., et al., 1987, A\&AS 67, 447

\bibitem[1993]{Bat93}  Battistini P.L., B\`onoli F., Casavecchia M., et al., 1993, A\&A 272, 77

\bibitem[2000]{Bor00}  Borozdin K.N., Priedhorsky W.C., 2000, ApJ 542, L13

\bibitem[1993]{Bra93}  Braun R., Walterbos R.A.M., 1993, A\&AS 98, 327

\bibitem[1989]{Cap89}  Capaccioli M., Della Valle M., D'Onofrio M., et al., 1989, AJ 97, 1622

\bibitem[1990]{Col90}  Collura A., Reale F., Peres G., 1990, ApJ 356, 119

\bibitem[1984]{Cra84}  Crampton D., Cowley A.P., Hutchings J.B., Schade D.J., 
                       Speybroeck L.P.van, 1984, ApJ 284, 663

\bibitem[1988]{Cru88}  Cruddace R.G., Hasinger G.R., Schmitt J.H., 1988, 
                       The Application of a Maximum Likelihood Analysis to 
                       Detection of Sources in the Rosat Data base. 
                       In:  Murtagh F., Heck A. (eds.) Astronomy from Large 
                       Databases. ESO Conference and Workshop Proceedings 
                       No. 28, Garching/Germany, p. 177

\bibitem[1980]{Dod80}  D'Odorico S., Dopita M.A., Benevenuti P., 1980, A\&AS 40, 67

\bibitem[2000]{Gar00}  Garcia M.R., Murray S.S., Primini F.A., et al., 2000, ApJ 537, L23

\bibitem[1996]{Gre96}  Greiner J., Supper R., Magnier E.A., 1996, Supersoft X-ray Sources
                       in \object{M31}. In: Greiner J. (ed.) Supersoft X-ray Sources, Springer Verlag,
                       p. 75

\bibitem[1994]{Hai94}  Haiman Z, Magnier E.A., Lewin W.H.G., et al., 1994, A\&A, 286, 725

\bibitem[1992]{Has92}  Hasinger G., et al., 1992, GSFC OGIP Calibration Memo,
                       CAL/ROS/92-001

\bibitem[2000]{Imm00}  Immler S., 2000, ``The X-Ray Source Population of Nearby Spiral Galaxies'',
                       PhD thesis, Ludwig--Maximilians--Universit\"at M\"unchen, Germany

\bibitem[1999]{Irw99}  Irwin J.A., Bregman J.N, 1999, ApJ 527, 125

\bibitem[1972]{Jac72}  Jacchia L.G., 1972, in CIRA 1972: COSPAR International
                       Reference Atmosphere 1972, compiled by The Committee
                       for CIRA of Cospar Working Group 4. Akademie-Verlag,
                       Berlin, p. 227

\bibitem[1999]{Kah99}  Kahabka P., 1999, A\&A 344, 459

\bibitem[1976]{Kat76}  Kato T., 1976, ApJS 30, 397

\bibitem[1983]{Lon83}  Long K.S., Speybroeck L.P.van, 1983, X-Ray Emission 
                       from Normal Galaxies. In: Lewin W.H.G., 
                       Heuvel E.P.J.van den (eds.)
                       Accretion Driven Stellar X-Ray Sources.
                       Cambridge University Press, Cambridge, p. 117

\bibitem[1996]{Lon96}  Long K.S., Charles P.A., Blair W.P., et al., 1996,
                       ApJ 466, 750L

\bibitem[1992]{Mag92}  Magnier E.A., Lewin W.H.G., van Paradijs J., et al., 1992,
                       A\&AS 96, 379

\bibitem[1994a]{Mag94a} Magnier E.A., Lewin W.H.G., van Paradijs J., et al., 1994a, A\&AS,
                       submitted

\bibitem[1995]{Mag95}  Magnier E.A., Prins S., van Paradijs J., et al., 1995, A\&AS,
                       114, 215

\bibitem[1993]{Pri93}  Primini F.A., Forman W., Jones C., 1993, ApJ 410, 615

\bibitem[1997]{Rea97}  Read A.M., Ponman T.J., Stickland D.K., 1997, MNRAS 286, 626

\bibitem[1991]{Sha91}  Sharov A.S., Alksnis A., 1991, Ap\&SS 180, 273

\bibitem[1992]{Sha92}  Sharov A.S., Alksnis A., 1992, Ap\&SS 190, 119

\bibitem[2001]{Shi01}  Shirey R., Soria R., Borozdin K., et al., 2001, A\&AL, 
                       365, 195

\bibitem[1994]{Shu94}  Shulman E., Bregman J.N., 1994, ROSAT HRI Observations of M33.
                       In: Schlegel E.M., Robert P. (eds.) The Soft X-ray Cosmos.
                       AIP Conference Proceedings \#313, AIP, New York, p. 345

\bibitem[1992]{Sno92}  Snowden S.L., Plucinsky P.P., Briel U., Hasinger G., 
                       Pfeffermann E., 1992, ApJ 393, 819

\bibitem[1993]{Sno93}  Snowden S.L., Freyberg M.J., 1993, ApJ 404, 403

\bibitem[1979]{Spe79}  van Speybroeck L.P., Epstein A., Forman W., et al., 
                       1979, ApJ 234, L45

\bibitem[1981]{Spe81}  van Speybroeck L.P., Bechtold J., 1981, X-Ray Emission 
                       from Normal Galaxies. In: Giacconi R. (ed.) 
                       X-Ray Astronomy with the Einstein Satellite. 
                       Reidel, Dordrecht, p. 153

\bibitem[1997]{Sup97}  Supper R., Hasinger G., Pietsch W., et al., 1997, A\&A 317,
                       328 [S97]

\bibitem[1991]{Tri91}  Trinchieri G., Fabbiano G., 1991, ApJ 382, 82 [TF]

\bibitem[1999]{Tri99}  Trinchieri G., Israel G.L., Chippetti L., et al., 1999, 
                       A\&A 348, 43 

\bibitem[1999a]{Vog99a} Vogler A., Pietsch W., 1999a, A\&A 342, 101

\bibitem[1999b]{Vog99b} Vogler A., Pietsch W., 1999b, A\&A 352, 64

\bibitem[1993]{Zim93}  Zimmermann H.U., Belloni T., Izzo C., Kahabka P., 
                       Schwentker O., 1993, MPE Report 244, Ed. 3


\end{thebibliography}
\end{document}